\documentclass[showpacs,twocolumn,aps,prx,floatfix,superscriptaddress,noshowpacs,longbibliography]{revtex4-1}
\usepackage{color}
\usepackage{psfrag}

 \usepackage[normalem]{ulem}
\usepackage{amsthm}
\usepackage{amsmath}
\usepackage{amsfonts}
\usepackage{amssymb}

\usepackage{braket}

\usepackage[utf8]{inputenc}
\usepackage{t1enc}
\usepackage[mathlines]{lineno}

\usepackage{amsthm,amsmath,amssymb,graphicx,bm,color,mathrsfs,verbatim,epstopdf,dcolumn,dsfont}
\usepackage{comment}

\graphicspath{{figs/},{./}}

\usepackage{hyperref}
\hypersetup{ 
	colorlinks   = true
}

\newcommand{\be}{\begin{equation}}
\newcommand{\ee}{\end{equation}}

\begin{document}


\title{Geometric Speed Limit of Accessible Many-Body State Preparation}
\author{Marin Bukov}
\email{mgbukov@berkeley.edu}
\affiliation{Department of Physics, University of California, Berkeley, CA 94720, USA}

\author{Dries Sels}
\affiliation{Department of Physics, Boston University, 590 Commonwealth Ave., Boston, MA 02215, USA}
\affiliation{Department of Physics, Harvard University, 17 Oxford St., Cambridge, MA 02138, USA}
\affiliation{Theory of quantum and complex systems, Universiteit Antwerpen, B-2610 Antwerpen, Belgium}

\author{Anatoli Polkovnikov}
\affiliation{Department of Physics, Boston University, 590 Commonwealth Ave., Boston, MA 02215, USA}

\begin{abstract}
We analyze state preparation within a restricted space of local control parameters between adiabatically connected states of control Hamiltonians. We formulate a conjecture that the time integral of energy fluctuations over the protocol duration is bounded from below by the geodesic length set by the quantum geometric tensor. The conjecture implies a geometric lower bound for the quantum speed limit (QSL). We prove the conjecture for arbitrary, sufficiently slow protocols using adiabatic perturbation theory and show that the bound is saturated by geodesic protocols, which keep the energy variance constant along the trajectory. Our conjecture implies that any optimal unit-fidelity protocol, even those that drive the system far from equilibrium, are fundamentally constrained by the quantum geometry of adiabatic evolution. When the control space includes all possible couplings, spanning the full Hilbert space, we recover the well-known Mandelstam-Tamm bound. However, using only accessible local controls to anneal in complex models such as glasses or to target individual excited states in quantum chaotic systems, the geometric bound for the quantum speed limit can be exponentially large in the system size due to a diverging geodesic length. We validate our conjecture both analytically by constructing counter-diabatic and fast-forward protocols for a three-level system, and numerically in nonintegrable spin chains and a nonlocal SYK model.
\end{abstract} 

\date{\today}
\maketitle

\section{\label{sec:intro}Introduction}

The Quantum Speed Limit (QSL) is the minimum time, $T_\mathrm{QSL}$, required to prepare a quantum state with unit fidelity. Understanding the physics behind it is anticipated to lead to significant advances in the field of quantum computing~\cite{nielsen}, which is based to a large extent on the ability to reliably manipulate the population of quantum states. The quantum speed limit  is also of prime importance for experimental quantum emulators, such as cold atoms~\cite{bason_12,vanfrank_16,wigley_16}, trapped ions~\cite{islam_11,senko_15,jurcevic_14}, and superconducting qubits~\cite{barends_16}, which require preparing quantum states with high fidelity before they can be studied. The origin of its physical meaning is rooted deeply in the Heisenberg energy-time uncertainty principle~\cite{deffner_17}, which implies that the time over which a quantum process occurs is intimately tied to the energy uncertainty $\Delta E$ it leads to. This was recognised by Mandelstam and Tamm~\cite{mandelstam_45,fleming_73,bhattacharyya_83,anandan_90}, who used it to introduce the lower bound $T_\mathrm{QSL}\geq \hbar \pi/(2\Delta E)$.

In recent years quantum speed limits have been studied ever more extensively, and various improved bounds and alternative derivations have been proposed~\cite{levitin_09,jones_10,uffink_93,margolus_98,taddei2013quantum,pires2016generalized}, including generalizations to mixed states~\cite{campaioli_17} and open systems~\cite{delCampo2013quantum}. In particular, it has been noticed that the bound can be sharpened by the \emph{absolute} geodesic length $\mathcal{L} = \arccos\vert\langle\psi_i|\psi_\ast\rangle\vert$~\footnote{whenever the states are not orthogonal}, leading to
\begin{equation}
T_\mathrm{QSL}\geq\hbar\frac{\arccos\vert\langle\psi_i|\psi_\ast\rangle\vert}{\Delta E},
\label{eq:QSL_1}
\end{equation}
for an initial state $|\psi_i\rangle$ and a target state $|\psi_\ast\rangle$. Unfortunately, this bound is of limited practical use in quantum many-body systems, where $\Delta E\sim \sqrt{L}$ scales with the system size $L$, and hence in the thermodynamic limit the bound becomes trivially $T_\mathrm{QSL}\geq 0$, misleadingly suggesting that it is possible to prepare any many-body state in no time. 

It is not hard to see that this issue arises due to the lack of constraints on the allowed terms in the Hamiltonian used to prepare the target state~\cite{campaioli2017enhancing}. In other words, since the bounds are based on generic geometric arguments, they must hold for any Hamiltonian. However, if one can fine tune the Hamiltonian arbitrarily, the quantum brachistochrone problem becomes almost trivial to solve~\cite{carlini2006}. In fact, the bound is tight, because the equality holds when the Hamiltonian is unconstrained: performing $H_\ast=i(|\psi_\ast\rangle \langle \psi_i | - |\psi_i\rangle \langle \psi_\ast|)/\sqrt{2}$, effectively realizes the $\sigma^y$ Pauli operator between the initial and the target states, saturating the bound. While this is admittedly not a problem in simple setups, such as a two-level system, where the control space is sufficiently small, it quickly becomes the bottleneck for many-body Hamiltonians, in which the realization of nonlocal terms like $H_\ast$ requires access to exponentially many couplings, and exponential sensitivity to fine-tune them. Indeed, realizable protocols only control local physical couplings and require much longer times, such that the bound~\eqref{eq:QSL_1} becomes useless. It does not tell us anything about how it is to prepare the target state. 

Let us illustrate this point explicitly. Consider a system of $L$ noninteracting qubits, prepared in some product initial state $|\psi_i\rangle=|\!\downarrow\cdots\downarrow\rangle$ and subject to a Hamiltonian $H=\sum_i H_i$. We want to transfer the population into the target product state $|\psi_*\rangle=|\!\uparrow\cdots\uparrow\rangle$. On the single-qubit level, it is optimal to do a $\pi$-pulse around the $y$- (or $x$-) axis, i.e.~$H_i=\Delta \sigma^y_i$, such that $T_{\rm QSL}^{(L=1)}=\hbar\pi/(2\Delta)$. Clearly, the existence of $L$ independent qubits does not make the process any faster. On the other hand, the energy fluctuations in the total system are $\Delta E=\Delta \sqrt{L}$, so the expression~\eqref{eq:QSL_1} suggests that it would be possible to rotate the spins faster. This fallacious argument shows how the standard quantum speed limit bounds are based on the premise that one can access the full Hilbert space to construct the optimal driving Hamiltonian. In the present example, this bound will be achievable only if one can realize the Hamiltonian $H_\ast \propto i (|\!\uparrow\cdots\uparrow\rangle\langle \downarrow\cdots\downarrow \!| - \mathrm{h.c.})/\sqrt{2}$ which transfers the population from the initial into the target state by rotating it into a macroscopic Schr\"odinger cat (GHZ) state at intermediate times. In experiments, where one only has \emph{local} control over the system, one simply cannot implement this evolution. Moreover, in more complex interacting setups the structure of the target state itself is very complicated so $H_\ast$ will not only be non-local but exponentially complex. One intuitively expects that $T_\mathrm{QSL}^{(L)}$ should generically increase with $L$ as it is usually much harder to prepare many-body states with a good fidelity, especially in complex systems.

Quantum state preparation has enjoyed renewed attention from the theoretical community in the last decade. Analytically, ideas known as Shortcuts to Adiabaticity have been put forward, developing the concepts of counter-diabatic (CD) and fast-forward (FF) driving protocols~\cite{vitanov_96,demirplak_03,demirplak_05,demirplak_08,berry_09,masuda_09,jarzynski_13,torrontegui_13,delcampo_13,deffner_14,kolodrubetz_16,sels_16,baksic_16,patra_17,jarzynski_17,bravetti_17,agarwal2017fast,stefanatos2018maximizing,hartmann2018rapid}. counter-diabatic driving studies the engineering of time-dependent counter-diabatic Hamiltonians, which generate transitionless time evolution [in the instantaneous basis of the original Hamiltonian] far away from the adiabatic limit. Also away from the adiabatic limit but allowing to create excitations during the evolution, fast-forward Hamiltonians are designed to steer the system into the  target state in a fixed amount of time. In the mean time, numerically, the state preparation paradigm has been formulated as an optimisation problem~\cite{rabitz_98,glaser_98,lloyd_14,glaser_15,bukov_17RL,bukov_17symmbreak,day_17,sorensen_18,ho2018efficient}. Recently, stochastic descent, gradient-based GRAPE~\cite{grape_05} and CRAB~\cite{caneva_11}, and model-free Machine Learning~\cite{ML_review,judson_92,chen_14,chen_14_ML,bukov_17RL,yang_17,dunjko_17,august_18,foesel_18,zhang2018automatic,niu2018universal,albarran2018measurement,bukov2018reinforcement} have proven useful algorithms to find approximate fast-forward Hamiltonians in single-particle and many-body systems.

In this work, we formulate a conjecture and give numerical and analytical evidence supporting the validity of a new, \emph{geometric} lower bound on the quantum speed limit  (cf.~Eq.~\eqref{eq:conjecture}, \eqref{eq:bound} below). This bound implies that the quantum speed limit  is controlled by the geodesic length between the initial and the target state \emph{in the eigenstate manifold} set by the control parameter space. Based on this conjecture, we show  that the adiabatic limit and the associated quantum geometry~\cite{kolodrubetz_16} constrain the time of possible unit-fidelity protocols both in single-particle and complex many-body systems. From our conjecture it also follows that the quantum speed limit  for all protocols is bounded by the quantum speed limit  for counter-diabatic protocols, which generally cannot be implemented within the constrained control parameter space, but for which the geodesic bound can be rigorously proven using recent results from Ref.~\cite{funo_17}. 

\section{\label{sec:conjecture}Geometric Bound Conjecture}

Consider a system described by the Hamiltonian $H(\lambda)$, where $\lambda$ is the control parameter which couples to a local operator. To simplify the discussion we assume that the control parameter has a single component~\footnote{In the end of this section, we comment how results of this work can be extended to the multi-component case $H(\vec\lambda)$.}. At time $t=0$ we prepare the ground state (GS) $\vert\psi(t=0)\rangle=\vert\psi_0(\lambda_i)\rangle$. We want to transfer the population with unit probability over a \emph{finite} time span $T$ from this initial state into a target state $\vert\psi(t\!=\!T)\rangle=\vert\psi_0(\lambda_\ast)\rangle$, which (up to an overall phase) is the ground state of $H(\lambda_\ast)$~\footnote{Instead of ground states we can consider any other pair of adiabatically connected states as we demonstrate in Sec.~\ref{subsec:excited_states}.}. In order to implement such a protocol we only allow Hamiltonians of the form $H(t)\equiv H(\lambda(t))$, which depend on time solely through the control function $\lambda(t)$. Such constrained Hamiltonians, if they prepare the target state with unit fidelity, are called fast-forward Hamiltonians: $H_\mathrm{FF}(t)\equiv H_\mathrm {FF}(\lambda((t))$~\footnote{In our setup we want to avoid rather special situations, in which the instantaneous eigenstate $\ket{\psi_0(\lambda)}$ of $H(\lambda)$ is a periodic function of $\lambda$, such that $\ket{\psi_0(\lambda_\ast)}=\ket{\psi_0(\lambda_i)}$ and the control problem is trivial. Hence, we additionally assume that $\braket{\langle \psi_0(\lambda) | \psi_0(\lambda_i) }$ is a monotonically decreasing function of $\lambda-\lambda_i$.}.

Whenever preparing the target state with unit probability (or unit fidelity) is possible, the system is called controllable. By the adiabatic theorem, for any non-degenerate Hamiltonian $H(\lambda)$ the problem becomes asymptotically controllable in the limit $T\to\infty$. Notice that, in general, there may exist multiple protocols which yield unit fidelity. Any unit-fidelity protocol obtained using Optimal Control methods gives rise by definition to a fast-forward Hamiltonian.

\emph{\bf Conjecture.---}Let us formulate the following conjecture: for any fast-forward Hamiltonian $H_\mathrm{FF}(\lambda(t))$ the energy fluctuations, averaged over the protocol duration, are larger than the geodesic length $\ell_\lambda$:
\be
\int_0^T \mathrm{d}t \sqrt{\delta E_{FF}^2(t)}\equiv\ell_t
\geq \ell_\lambda\equiv \int_{\lambda_i}^{\lambda_\ast} \mathrm{d}\lambda \sqrt{g_{\lambda\lambda}},
\label{eq:conjecture}
\ee
where the parameter $\lambda$ changes along a fixed unit-fidelity protocol in an arbitrary way. 
Note that we define the geodesic length $\ell_\lambda$ within the control space; it is generally larger than the distance between wavefunctions (i.e.~the absolute geodesic).  In particular, for extensive systems with $L$ degrees of freedom and local controls, the RHS of Eq.~\eqref{eq:conjecture} typically scales as $\sqrt{L}$ while the distance between wavefunctions is always bounded from above by $\pi/2$.
Further, in Eq.~\eqref{eq:conjecture},
\begin{eqnarray}
&&\delta E_{FF}^2(t)\!=\! \langle \psi(t) | H_\mathrm{FF}^2(t) |\psi(t)\rangle_c\!=\!\langle \partial_t\psi(t) | \partial_t\psi(t)\rangle_c\!=\!g_{tt} \nonumber\\
&&g_{\lambda\lambda}\!=\!\langle\partial_\lambda\psi(t) | \partial_\lambda\psi(t)\rangle_c \!=\!\langle \psi_0 (\lambda) | \mathcal A_\lambda^2(\lambda)  |\psi_0(\lambda)\rangle_c
\label{eq:metric}
\end{eqnarray}
are the energy variance $\delta E_{FF}^2(t)$, which can be thought of as the time-time component of the geometric tensor $\delta E_{FF}^2(t)=g_{tt}$, and the eigenstate Fubini-Study metric tensor, $g_{\lambda\lambda}$, respectively. Here $|\psi_0(\lambda)\rangle$ is the instantaneous ground state of $H(\lambda)$, and $\mathcal A_\lambda$ is the adiabatic gauge potential~\cite{kolodrubetz_16}.  The ket $|\psi(t)\rangle$ denotes the time-evolved initial state under the Hamiltonian $H_\mathrm{FF}(t)$, which satisfies the boundary conditions $|\psi(0)\rangle=|\psi_0(\lambda_i)\rangle$  and $|\psi(T)\rangle=|\psi_0(\lambda_\ast)\rangle$. We emphasize the difference between the evolved and the instantaneous states: $|\psi(t)\rangle\neq |\psi_0(\lambda(t))\rangle$. The subscript $_c$ denotes the connected expectation value: $\langle H_\mathrm{FF}^2 \rangle_c = \langle H_\mathrm{FF}^2\rangle - \langle H_\mathrm{FF} \rangle^2$.  

To motivate the conjecture, notice that this bound is tight and can be saturated in the adiabatic limit. Indeed, from Adiabatic Perturbation Theory (APT) it follows that~\cite{kolodrubetz_13, kolodrubetz_16} 
\be
\label{eq:deltaE}
\delta E_{FF}^2=\langle \psi(t) | H_\mathrm{FF}^2(t) |\psi(t)\rangle_c = \dot \lambda^2 g_{\lambda\lambda} + \mathcal{O}(\dot{\lambda}^4). 
\ee
Hence, for any monotonic $\lambda(t)$ the bound~\eqref{eq:conjecture} is saturated in the adiabatic limit. Moreover, it is easy to see that at least for any real-valued Hamiltonian, satisfying instantaneous time-reversal symmetry, the next-order correction to Eq.~\eqref{eq:deltaE} scales as $\dot \lambda^4$ with a non-negative pre-factor, such that $g_{tt}-\dot \lambda^2 g_{\lambda\lambda}\geq 0$. This fact follows immediately from the structure of APT where all the coefficients in the expansion of the wave function in the instantaneous basis in powers of $\dot\lambda$ are imaginary in linear order, and real-valued in quadratic order [see Eq.~(12) in Ref.~\cite{degrandi_13}]:
\[
|\psi(t)\rangle=|\psi_0\rangle+i \dot\lambda |\psi^{(1)}\rangle+\dot\lambda^2|\psi^{(2)}\rangle,
\]
where $|\psi_0\rangle$,  $|\psi^{(1)}\rangle$ and $|\psi^{(2)}\rangle$ are real-valued functions. This observation, in turn, implies that there is no interference between the $\dot\lambda$ and $\dot\lambda^2$ contributions to the energy variance. In particular there is no $\dot\lambda^3$ contribution, and hence the quadratic and quartic terms above come from squares and are non-negative:
\[
\delta E^2=\dot\lambda^2 \langle \psi^{(1)} | H^2 |\psi^{(1)}\rangle_c+\dot\lambda^4 \langle \psi^{(2)} | H^2 |\psi^{(2)}\rangle_c+O(\dot\lambda^6).
\] 
Therefore, at least perturbatively, the bound is satisfied for any sufficiently slow protocol. We note that within APT, the $\dot\lambda^4$ contribution is treated on the same footing as the squared acceleration term $\ddot\lambda^2$ because $\mathrm d_t (\dot\lambda)=\dot\lambda \partial_{\lambda} (\dot \lambda)\sim \dot\lambda^2$. Indeed the linear in acceleration correction to the wave function also becomes imaginary~\cite{degrandi_13, kolodrubetz_16}.

Even though we formulated the bound for fast-forward Hamiltonians, the conjecture is intimately related to CD driving protocols. In a recent work Funo et al.~derived that, for any counter-diabatic protocol with monotonic $\lambda(t)$ the inequality Eq.~\eqref{eq:conjecture} is always saturated~\cite{funo_17}. This can be seen as follows: using the counter-diabatic Hamiltonian,
\[
H_{\rm CD}=H(\lambda(t))+\dot\lambda \mathcal A_\lambda,
\]
the system follows the instantaneous ground state of $H(\lambda)$: $|\psi(t)\rangle=|\psi_0(\lambda(t)\rangle$. Then evaluating the variance of $H_{\rm CD}$ one can convince oneself that the only non-zero contribution comes from the gauge potential term: 
\[
\langle \psi(t) |H_{\rm CD}^2|\psi(t)\rangle_c=\dot \lambda^2 \langle \psi_0 | \mathcal A_\lambda^2 |\psi_0\rangle_c=\dot\lambda^2 g_{\lambda\lambda}
\]
and hence $g_{tt}=\dot\lambda^2 g_{\lambda\lambda}$. This leads to the interesting observation that the leading non-adiabatic contribution to the energy variance without CD driving, is identical to the energy variance coming from the gauge potential in the CD protocols. However, a major difference is that for counter-diabatic protocols this result applies to arbitrarily fast protocols where APT does not hold. 

We point out that CD protocols usually require adding new control parameters, e.g.~for any real-valued Hamiltonian $H(\lambda(t))$ the gauge potential is imaginary so that any counter-diabatic protocol necessarily breaks instantaneous time-reversal symmetry. Moreover gauge potentials for generic Hamiltonians are highly fine-tuned typically requiring hard-to-implement \emph{non-local} operators. In certain simple cases it is possible to explicitly map counter-diabatic protocols to fast-forward protocols by an extra unitary rotation~\cite{kolodrubetz_16} but in general this unitary is hard to find.

If correct, the conjecture has immediate far-reaching implications: 
\begin{itemize}
	
	\item[(i)] \emph{Minimum Time Bound}: using the Cauchy-Schwarz inequality, we have
	\begin{eqnarray}
	&&\left(\int_0^T \mathrm{d}t \sqrt{\langle \psi(t) | H_\mathrm{FF}^2(t) |\psi(t)\rangle_c}\right)^2 \nonumber\\
	&&\leq T \int_0^T \mathrm{d}t \langle \psi(t) | H_\mathrm{FF}^2(t) |\psi(t)\rangle_c=T^2 \overline{\delta E_{FF}^2},
	\end{eqnarray}
	where $\overline{\delta E_{FF}^2}$ is the time average energy variance over the protocol duration $T$. Combining this result with the conjecture, and setting $T=T_\mathrm{QSL}$ to be the minimum time required to prepare the target state with unit fidelity, we obtain the following bound
	\be
	T_\mathrm{QSL}\geq {\ell_\lambda\over \sqrt{\overline{\delta E_{FF}^2}}},
	\label{eq:bound}
	\ee
	which holds for any optimal protocol. This bound is tight because it is saturated for slow geodesic protocols~\cite{tomka_16}. For these protocols the inequality~\eqref{eq:conjecture} is saturated by the validity of APT. In addition, in geodesic protocols the energy variance is kept constant along the trajectory $\dot\lambda^2 g_{\lambda\lambda}={\rm const}_t$, which sets the velocity profile.  In this case the Cauchy-Schwarts inequality becomes an equality and hence the bound~\eqref{eq:bound} is saturated.
		
	\item[(ii)] \emph{Local control between eigenstates is exponentially slow for systems satisfying the Eigenstate Thermalization Hypothesis:} note that the metric tensor can be expressed through the non-equal time correlation function~\cite{degrandi_13,kolodrubetz_13}:
	\begin{eqnarray}
	g_{\lambda\lambda}&=&-\Re \int_{0}^\infty \mathrm{d}t\,t \,\langle \psi_0 | M_\lambda(t) M_\lambda(0)|\psi_0\rangle\nonumber\\
	&=&\sum_{n\neq 0} {|\langle n |M_\lambda| 0\rangle|^2\over (E_n-E_0)^2}
	\end{eqnarray}
	where $M_\lambda(t)=-\partial_\lambda H(t)$ is the conjugate force with respect to the parameter $\lambda$ in the Heisenberg representation~\cite{kolodrubetz_16}. If we target ground states of systems with glassy dynamics or exact many-body excited states in generic systems satisfying the eigenstate thermalization hypothesis (ETH)~\cite{ETH_review}, then the geodesic length $\ell_\lambda$ scales exponentially with the system size $L$, while the energy variance is at most extensive. Therefore, the conjecture implies that at best the fast-forward Hamiltonian with local control can reach the target state only at exponentially long times. Interestingly, according to this bound, isolated critical points can be crossed at non-extensive times, which can be seen as follows. The geodesic length scales as $\sqrt{L}$ for any phase transition with the correlation length exponent $\nu<1$~\cite{kolodrubetz_13}, and so does the energy variance (if we drive the system with some global coupling); therefore, the ratio in Eq.~\eqref{eq:bound} is system-size independent. Intuitively, such finite-time protocols can be e.g.~realized by driving the system fast everywhere except near the critical point~\cite{barankov_08, tomka_16}.

	\item[(iii)] \emph{Generalization to multi-parameter drives:} our results immediately generalize to systems with a multi-component parameter space $\vec \lambda$. Then by $\ell_\lambda$ in Eq.~\eqref{eq:bound} one understands the geodesic length, which is defined as the minimum over all accessible paths connecting $\vec \lambda_i$ and $\vec \lambda_\ast$.
	
	\item[(iv)] \emph{The conjecture only applies to unit fidelity protocols:} It is interesting to see if and how the conjecture can be extended to protocols which require unit fidelity with some non-zero tolerance factor.
	
	\item[(v)] \emph{The conjecture gives a bound, which generally survives the classical limit $\hbar\to 0$} since both sides of Eq.~\eqref{eq:conjecture} represent well-defined quantities in the classical limit~\cite{kolodrubetz_16, sels_16,funo_17}. The same applies to the inequality~\eqref{eq:bound} bounding the speed limit. Note that with $\hbar$ explicitly included into the equations, $\ell_\lambda=\int d\lambda \sqrt{\hbar^2 g_{\lambda\lambda}}$ and it is the product $\hbar^2 g_{\lambda\lambda}$ which is well-defined in the classical limit~\cite{kolodrubetz_16}.
	
\end{itemize} 

Despite its plausibility, a direct proof of this conjecture has so far remained elusive due to the absence of a general procedure to obtain fast-forward Hamiltonians analytically. In the following, we demonstrate its validity beyond APT in a variety of systems of increasing complexity ranging from few-spin models to a non-integrable Ising chains: (i) analytically, using specific exactly solvable examples, showing a proof-of-concept strategy to derive fast-forward Hamiltonians by unitarily rotating counter-diabatic protocols, and (ii) numerically, using Optimal Control algorithms.

\section{\label{sec:analytical}Analytical Verification of the Geometric Bound Conjecture}

In this section we consider two exactly-solvable examples to analytically verify the validity of the conjecture. To this end, we first show how one can use counter-diabatic driving to find a fast-forward Hamiltonian. The first example will be a two-level system for which the conjecture reduces to the original Mandelstam-Tamm bound. We nevertheless want to show the proof as it highlights how going from a counter-diabatic to a fast-forward protocol increases the time length and hence the QSL. The second example is a three-level system, where the conjecture becomes much less trivial and gives a larger value of quantum speed limit  than the Mandelstam-Tamm bound.

\subsection{\label{subsec:2LS}Two-Level System}

Consider first  the prototypical model of a two-level system (2LS) governed by the following Hamiltonian:
\begin{equation}
\label{eq:H_2LS}
H_\mathrm{2LS}(t) = -g S^z - \lambda(t)S^x,
\end{equation}
where $g$ is a fixed magnetic field along the $z$-axis and $\lambda(t)$ is an a priori unknown optimal protocol. 
We prepare the system in the ground state $|\psi_i\rangle$ of $H_\mathrm{2LS}(\lambda_i\!=\!-2g)$ and seek a function $\lambda(t)$ which targets the ground state $|\psi_\ast\rangle$ at $\lambda_\ast\!=\!+2g$ in time $T$, following unitary evolution under $H_\mathrm{2LS}(t)$. State preparation in this model has been discussed extensively in the context of various approaches, and analytical expressions for the optimal protocols have been derived~\cite{hegerfeldt_13}. As we mentioned we will use this example to highlight connections between counter-diabatic and fast-forward protocols.

Before we dive into this analysis, notice a quick but curious fact: the initial and target states are related by the rotation $|\psi_\ast\rangle = \exp(-i\pi S^z)|\psi_i\rangle$. Hence, the static Hamiltonian $H_\mathrm{FF}(t) = -g S^z$ is a legitimate fast-forward Hamiltonian for $T=\pi$ (with $\lambda(t)\equiv 0)$. Let us compute the left-hand side (LHS) and the right-hand side (RHS) in Eq.~\eqref{eq:conjecture} separately. On the RHS, note that the geodesic length is $\ell_\lambda=\theta$, where $\tan\theta=\lambda_i/g$. On the LHS, on the other hand, we have $\langle\psi(t) |(S^z)^2|\psi(t)\rangle\!=\!1/4$ and $\langle\psi(t) |S^z|\psi(t)\rangle \!=\! \cos(\theta)/2$, and hence $\ell_t \!=\! \sin(\theta)/2$. Therefore, the inequality~\eqref{eq:conjecture} yields $\pi\sin\theta\geq 2\theta$, which is indeed always true for $\theta<\pi/2$, and hence the conjecture holds true for this special case.

The counter-diabatic protocol amounts to adding an extra (counter) term to the Hamiltonian which keeps it in the instantaneous ground state~\cite{demirplak_03,berry_09,jarzynski_13,delcampo_13, kolodrubetz_16}:
\begin{equation}
H_\mathrm{CD}(t)\!=\!H(t)+\dot \lambda(t)\mathcal{A}_\lambda(t),
\label{eq:H_CD}
\end{equation}
where 
\[
\mathcal A_\lambda={g\over \lambda^2+g^2} S^y
\]
is the (adiabatic) gauge potential with respect to the parameter $\lambda$ (see e.g.~Ref.~\cite{kolodrubetz_16} for details). However, the counter-diabatic protocol kicks the Hamiltonian out of the original control space by adding a magnetic field along $y$-direction. In order to map the counter-diabatic protocol to a valid fast-forward protocol, we need to perform an additional unitary rotation, as was first discussed in Ref.~\cite{masuda_09}:
\begin{equation}
H_\mathrm{FF}(t)=R^\dagger(t)H_\mathrm{CD}(t)R(t)-iR^\dagger(t)\partial_t R(t)\sim H(t),
\end{equation}
where $R(t)$ is a unitary change-of-frame matrix, which is equal to the identity in the beginning and in the end of the protocol: $R(0)\!=\!\hat 1\!=\!R(T)$. In this case it is easy to see that the wave function $|\psi(t)\rangle$ follows the ground state of a gauge equivalent Hamiltonian $H'(t)=R^\dagger H(\lambda(t)) R$. Therefore $|\psi(t)\rangle$ coincides with the initial and target states in the beginning and in the end of the protocol.

Let us now take the extreme case of the fastest counter-diabatic protocol $\dot\lambda\to\infty$, where the counter-diabatic Hamiltonian reduces to the rate times the gauge potential (the calculation away from the infinite-speed limit is shown in App.~\ref{sec:2LS_full}):
\[
H_{\rm CD}=\dot\lambda{g\over \lambda^2+g^2} S^y
\]
For the unitary $R(t)$ we can choose
\begin{equation}
R(t)=\exp\left(-i\frac{\pi}{2}\left[\Theta(t)+\Theta(T-t)\right]S^x\right ),
\end{equation} 
where $\Theta(t)$ is the Heaviside step function. This transformation rotates $S^y$ to $S^z$. Note that $R(t)$ is constant except at $t=0,T$ giving rise to the pulse-like contributions from $R^\dagger(t)\partial_t R(t)$ to the fast-forward Hamiltonian:
\be
\label{eq:HFF_2LS}
H_\mathrm{FF}(t)=-\dot \lambda {g\over g^2 +\lambda^2} S^z + \frac{\pi}{2}\left[\delta(t)-\delta(T-t)\right]S^x
\ee
with $\delta(t)\!=\!\partial_t\Theta(t)$ the Dirac delta function. Finally, to make the $z$-magnetic field time independent, we can rescale the time according to
\[
\mathrm dt'=\mathrm d t {\dot \lambda \over g^2+\lambda^2}= {\mathrm d\lambda \over g^2+\lambda^2}.
\]
Then, using that $\delta(t)=\delta(t')|\mathrm dt'/\mathrm dt|$, we find
\be
H_\mathrm{FF}(t')=-g S^z + \frac{\pi}{2}\left[\delta(t')-\delta(T'-t')\right]S^x.
\ee
The total protocol time $T'\equiv T_{\rm QSL}$, which sets the quantum speed limit  in this case~\cite{hegerfeldt_13}, can be found as
\begin{multline}
\label{eq:T_QSL_2LS}
T_\mathrm{QSL}=\int_0^{T_\mathrm{QSL}}\mathrm{d}t'=\int_{\lambda_i}^{\lambda_f}\mathrm  d\lambda {dt'\over d\lambda}= \int_{\lambda_i}^{\lambda_\ast}\mathrm{d}\lambda {1\over g^2+\lambda^2}=\\
\frac{1}{g}\left[\arctan\left(\frac{g}{\lambda_\ast}\right) - \arctan\left(\frac{g}{\lambda_i}\right)\right]={2\theta\over g}.
\end{multline}

Let us now check the conjecture for this quantum speed limit  protocol. To evaluate the LHS of Eq.~\eqref{eq:conjecture}, notice first that both $\delta$-function kicks can be interpreted as a free rotation under the Hamiltonian $H=S^x$ for the time $\pi/2$. Second, for a (piecewise) constant Hamiltonian the energy variance is (piecewise) conserved. Therefore, we need to add two contributions from the kicks and a contribution coming from the rotation around $z$-axis, leading to:
\begin{multline}
	\label{eq:2LS_LHS}
		\ell_t=\int_0^{T_{\rm QSL}}\mathrm{d}t' \sqrt{\langle\psi(t')|H^2_\mathrm{FF}(t')|\psi(t')\rangle_c} \\= {\pi\over 2}\sqrt{\langle\psi_i|\left(S^x\right)^2|\psi_i\rangle_c}+{\pi\over 2}\sqrt{\langle\psi_\ast|\left(S^x\right)^2|\psi_\ast\rangle_c}\\+T_{\rm QSL} \sqrt{\langle\psi(0_+)|\left(S^z\right)^2|\psi(0_+)\rangle_c},
\end{multline}
where $|\psi(0_+)\rangle$, is the wave function right after the first $\pi/2$ rotation around $x$-axis, which brings the spin to the $xy$-plane. Using that $\langle\psi(0_+)| (S^z)^2|\psi(0_+)\rangle_c = 1/4$ and $\langle\psi_i|(S^x)^2|\psi_i\rangle_c=\langle\psi_*|(S^x)^2|\psi_*\rangle_c \!=\! \cos^2(\theta)/4$ we find
\[
\ell_t = \theta + {\pi\over 2} \cos\theta.
\]
On the RHS of the conjecture~\eqref{eq:conjecture}, we have the geodesic length $\ell_\lambda=\int \mathrm d\lambda\sqrt{g_{\lambda\lambda}}$, where 
\begin{multline*}
g_{\lambda\lambda}=\langle \psi_0(\lambda) |\mathcal A_\lambda^2|\psi_0(\lambda)\rangle_c=\\
{g^2\over (\lambda^2+g^2)^2}\langle \psi_0(\lambda) |(S^y)^2|\psi_0(\lambda)\rangle_c={1\over 4}{g^2\over (\lambda^2+g^2)^2},
\end{multline*}
leading to $\ell_\lambda=\theta$ such that $\ell_t\geq \ell_\lambda$ is indeed satisfied. We see that in this simple example the difference between $\ell_t$ and $\ell_\lambda$ can be attributed to an extra rotation required to bring (at the QSL, kick) the $y$-gauge potential term back to the allowed $xz$-plane.

\subsection{\label{subsec:3LS_I}Three-Level System I}

With the exception of the two-level system example above and a few other free-particle systems~\cite{sels_16}, it is not known how to analytically compute the fast-forward Hamiltonian or the quantum speed limit $T_\mathrm{QSL}$ in more complicated systems. Below, we show that the ideas of mapping counter-diabatic to fast-forward driving protocols presented in Sec.~\ref{subsec:2LS}, can be used to identify other controllable models and compute the corresponding value for $T_\mathrm{QSL}$. Along the way, we unveil the difficulty and hidden complexity behind constructing fast-forward protocols in generic systems, and showcase a concrete example which features an intrinsic emerging dynamical gauge degree of freedom.

Consider the two-qubit system described by the Hamiltonian
\begin{equation}
H_\mathrm{3LS}(\lambda) = -2JS^z_1S^z_2 -g (S^z_1+S^z_2) - \lambda(S^x_1+S^x_2),
\label{eq:H_3LS}
\end{equation}
where, as before, $g$ and $\lambda$ are the magnetic field components along the $z$ and $x$-directions respectively, and $J=1$ is the $zz$-interaction strength which sets the reference energy scale. Let the initial and target states be the ground states of $H_\mathrm{3LS}(\lambda)$ for $\lambda_i=-2g=-\lambda_\ast$, respectively. Similar to Sec.~\ref{subsec:2LS}, our goal is to find a protocol $\lambda(t)$ which prepares the target state in time $T$, following evolution with the \emph{single-particle} Hamiltonian $H_\mathrm{2LS}(t)$. Due to the qubit-exchange symmetry of both $H_\mathrm{3LS}$ and $H_\mathrm{2LS}(t)$, the problem represents effectively a three-level system (3LS) with $SU(3)$ spanning the space of all possible observables. 

A priori, it is not clear whether such an optimal protocol exists, since the initial and target states are eigenstates of a fully interacting Hamiltonian, whereas during the evolution the system is non-interacting (decoupled). Note that, in general, this population transfer can only be achieved if and only if the entanglement entropy of each of the two qubits is the same in the initial and target states, as entanglement is preserved during evolution with the non-interacting Hamiltonian $H_\mathrm{2LS}$. This condition is clearly satisfied in our setup, since the states are related by the transformation $|\psi_\ast\rangle = \exp(-i\pi (S^z_1+S^z_2))|\psi_i\rangle$. Furthermore, the static Hamiltonian $H_\mathrm{FF}(t) = -g (S^z_1+S^z_2)$ is a legitimate fast-forward Hamiltonian for $T=\pi$, similar to the 2LS, c.f~Sec.~\ref{subsec:2LS}. It is straightforward to check that this fast-forward Hamiltonian satisfies the geometric bound conjecture~\eqref{eq:conjecture}.

Unfortunately, this only works for the protocol duration $T\!=\!\pi$, which immediately puts an upper bound on the QSL. For $T<\pi$, one can formally rely on general theorems in Optimal Control for systems on compact Lie groups~\cite{jurdjevic_72}, to argue the existence of a finite quantum speed limit  $T_\mathrm{QSL}>0$ for this problem. However, since the proofs are non-constructive, one cannot use them to directly check the validity of the geometric bound conjecture. Nonetheless, as we demonstrate now, one can apply the same strategy as the 2LS example in Sec.~\ref{subsec:2LS}. In particular, (i) we first compute the counter-diabatic Hamiltonian, and then (ii) use the latter to  derive a fast-forward protocol. However, in practice finding the correct frame transformation in step (ii) is a particularly difficult problem, since there is no straightforward way to identify the correct time-dependent rotation to map the interacting counter-diabatic Hamiltonian to the non-interacting $H_\mathrm{2LS}$. The procedure requires the use of non-commuting rotations in the $8$-dimensional operator manifold corresponding to the $SU(3)$ group which, due to their intrinsic time-dependence, lead to unwanted extra Galilean terms that take the transformed Hamiltonian outside the parameter manifold of $H_\mathrm{2LS}$. Moreover, an additional constraint is imposed by the boundary conditions imposing that the rotation reduces to the identity at $t=0,T$, c.f.~Sec.~\ref{subsec:2LS}. In the following, we demonstrate how to circumvent all these issues and construct a fast-forward Hamiltonian for the system in Eq.~\eqref{eq:H_3LS}.

Due to the small dimensionality of the Hilbert space, it is possible to find the exact adiabatic gauge potential in the ground state manifold. Note that, since the Hamiltonian $H_\mathrm{3LS}(\lambda)$ is real-valued, one can choose the gauge potential to be purely imaginary~\cite{kolodrubetz_16,sels_16}. There are only three linearly-independent imaginary matrices which can be shown to generate a $SU(2)\subset SU(3)$. Hence, in the most general form we have
\begin{equation}
\label{eq:A_3LS}
\mathcal{A}_\lambda\!=\!\alpha\left(S_1^y+S_2^y\right) + b\left(S_1^xS_2^y+S_1^yS_2^x\right) + \gamma\left(S_1^yS_2^z+S_1^zS_2^y\right),
\end{equation}
where $\alpha\!=\!\alpha(\lambda,b(\lambda))$ and $\gamma\!=\!\gamma(\lambda,b(\lambda))$ are fixed functions, which depend on the model parameters, and can be computed using, e.g., a variational principle~\cite{sels_16}. We leave details of such computation for the appendix App.~\ref{sec:controllability}. Let us only comment that because we are looking into the gauge potential for the ground state manifold, i.e.~the gauge potential which adiabatically evolves the ground state but allows mixing between the two excited states, the gauge potential is defined up to an emergent dynamical gauge degree of freedom $b\!=\!b(\lambda)$, which we use to our advantage in finding the fast-forward Hamiltonian.

Having computed the exact gauge potential, which governs the dynamics at the QSL, we now aim at finding a transformation $R(t)$ that brings the counter-diabatic Hamiltonian~\eqref{eq:A_3LS} to the original parameter manifold~\eqref{eq:H_2LS} with renormalized drive field and an overall time-dependent pre-factor. If we fix the dynamical gauge field $b(\lambda)$ to satisfy the following nonlinear differential equation
\begin{eqnarray}
&&b(\lambda)=2\partial_{\lambda}\arctan\left(\frac{\gamma(\lambda,b(\lambda))}{2\alpha(\lambda,b(\lambda))}\right), \nonumber\\
&& \gamma(\lambda_i,b(\lambda_i))\!=\!0\!=\!\gamma(\lambda_\ast,b(\lambda_\ast))
\end{eqnarray} 
one can show [see App.~\ref{sec:controllability}] that the non-abelian $SU(3)$-rotation
\begin{eqnarray}
R(t)\!&=&\!
\exp\left(-i\arctan\left(\frac{\gamma(t)}{2\alpha(t)}\right)\left(S_1^xS_2^y+S_1^yS_2^x\right)\right) \nonumber\\
\!&\times&\!\exp\left(-i\frac{\pi}{2}\left[\Theta(t)+\Theta(T-t)\right]\left(S^x_1+S^x_2\right)\right) 
\end{eqnarray}
obeys the boundary conditions $R(0)=1=R(T)$. Using this time-dependent transformation leads to the following fast-forward Hamiltonian at the QSL:
\begin{eqnarray}
\label{eq:HFF_3LS}
H_\mathrm{FF}(t)&=&\frac{\dot \lambda(t)}{2}\sqrt{4\alpha^2(t)+\gamma^2(t)}(S^z_1 + S^z_2) \nonumber\\
&&+ \frac{\pi}{2}\left[\delta(t)-\delta(t-T)\right]\left(S^x_1 + S^x_2\right).
\end{eqnarray}
We note in passing that the existence of $H_\mathrm{FF}$ is equivalent to a constructive proof of controllability, i.e.~a finite $T_\mathrm{QSL}\!<\!\infty$, since by definition all fast-forward Hamiltonians prepare the target state with unit fidelity. Thus, the above result establishes the relation between CD, fast-forward and Optimal Control for the problem of preparing interacting two-qubit states using a single-particle Hamiltonian. 

The above mapping works at the infinite-speed QSL, where $H_\mathrm{CD}=\dot\lambda\mathcal{A}_\lambda$. Unlike the 2LS discussed in Sec.~\ref{subsec:2LS} and App.~\ref{sec:2LS_full}, it is currently an open question how to construct the correct transformation away from the quantum speed limit  for this problem. 
\begin{figure}[t!]
	\includegraphics[width=1.0\columnwidth]{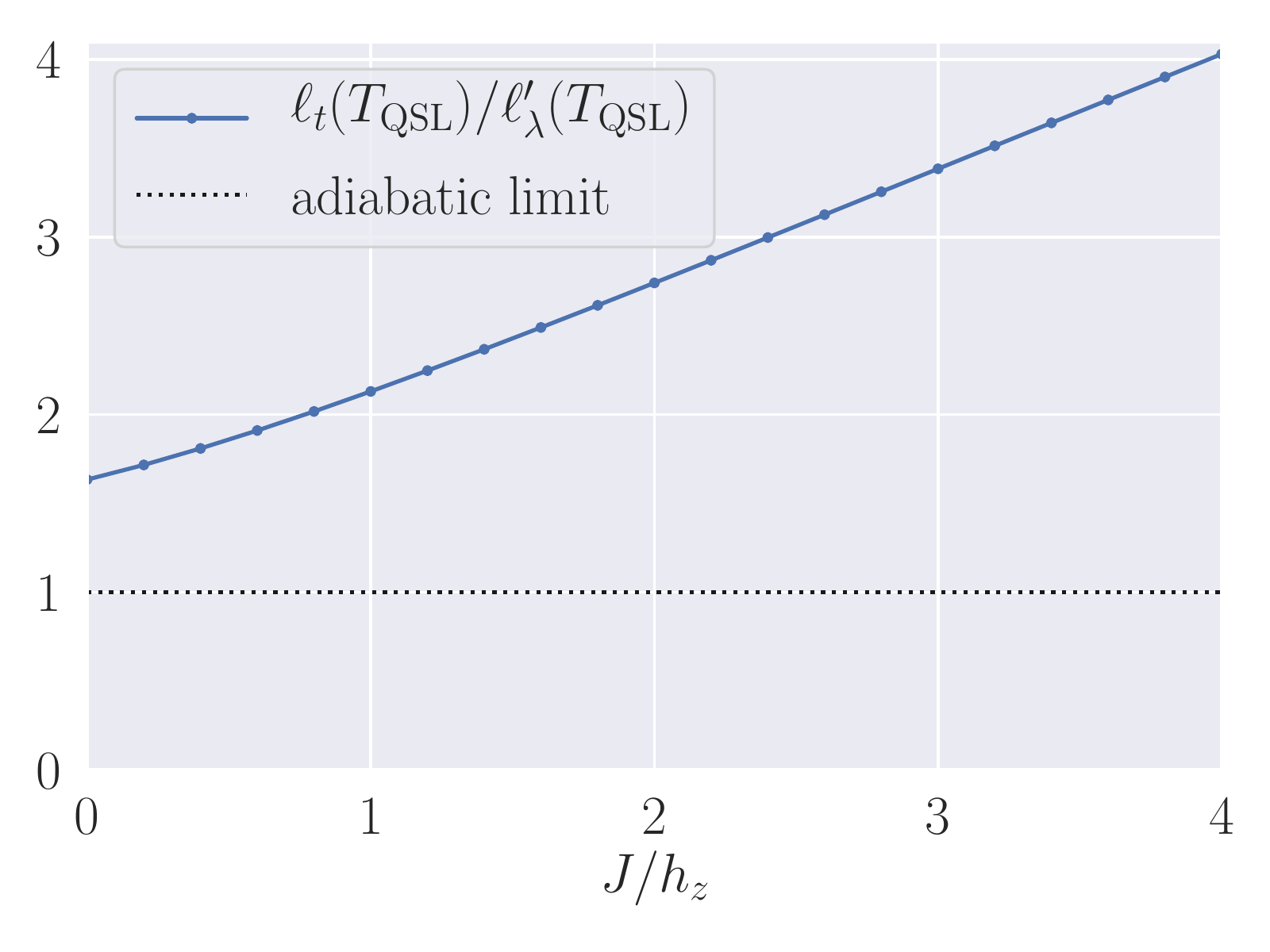}
	\caption{\label{fig:3LS_nonint} Numerical justification of the conjecture~\eqref{eq:conjecture}, at the quantum speed limit of preparing the interacting ground state of $H_\mathrm{3LS}$ with the non-interacting fast-forward Hamiltonian $H_\mathrm{2LS}(t)$. The parameters are $\lambda_i/g=-2=-\lambda_\ast/g$.}
\end{figure}
As in the 2LS example we can rescale the time as 
\be
\mathrm dt'={\mathrm d\lambda \over 2 g} \sqrt{4\alpha^2(t)+\gamma^2(t)}
\ee
such that the $z$-magnetic field is constant. Then, following the same strategy as in the 2LS, we obtain the expression for the QSL.
\begin{equation}
\label{eq:TQSL_3LS}
T_\mathrm{QSL}\!=\!\int_0^{T_\mathrm{QSL}}\mathrm{d}t'\!=\!\int_{\lambda_i}^{\lambda_\ast} \frac{\mathrm{d}\lambda}{2g}\sqrt{4\alpha^2(\lambda,b(\lambda))+\gamma^2(\lambda,b(\lambda))}.
\end{equation}

We can use the analytical results obtained above to verify the validity of the geometric bound conjecture~\eqref{eq:conjecture}. Once again, we shall compute the LHS and RHS separately. On the RHS, we need to compute the geodesic length $\ell_\lambda$. This requires some care for the current problem. Since we quench the interaction strength $J$ at $t=0,T$, so that for $0<t<T$ the time evolution remains free ($J=0$), we effectively have a two-parameter manifold $(\lambda,J)$. Thus, as we specified in Sec.~\ref{sec:conjecture}, point (iii), the geodesic length $\ell_\lambda$ on the RHS of~\eqref{eq:conjecture} is the minimum length $\ell_\lambda$ of all accessible paths connecting $(\lambda_i,J_i)$ and $(\lambda_*,J_*)$. An upper bound $\ell_\lambda'$ for this absolute minimum is given by the geodesic along $J=1$, which can easily be obtained from the gauge potential~\eqref{eq:A_3LS}. We emphasize that $\ell_\lambda'$ is independent of the choice for the dynamical gauge field $b(\lambda)$, as expected. Hence, to verify the conjecture, it suffices to show that $\ell_t\geq \ell_\lambda'$, since $\ell_\lambda'\geq \ell_\lambda$.

Let us now focus on the LHS and the number $\ell_t$. Notice that the calculation is formally equivalent to the one we carried out for the 2LS in Sec.~\ref{subsec:2LS}, due to the structure of the 3LS fast-forward Hamiltonian~\eqref{eq:HFF_3LS}. Thus, we just need to apply Eq.~\eqref{eq:T_QSL_2LS} using the fast-forward Hamiltonian~\eqref{eq:HFF_3LS} and the quantum speed limit  expression~\eqref{eq:TQSL_3LS}. Therefore, decomposing the LHS $\ell_t$ according to Eq.~\eqref{eq:2LS_LHS}, we arrive at
\begin{eqnarray}
\ell_t &=& \ell_\lambda' + {\pi\over 2}\sqrt{\langle\psi_i|\left(S^x\right)^2|\psi_i\rangle_c}+{\pi\over 2}\sqrt{\langle\psi_\ast|\left(S^x\right)^2|\psi_\ast\rangle_c} \nonumber\\
&\geq& \ell_\lambda',
\end{eqnarray}
where $\ell_\lambda'=T_{\rm QSL} \sqrt{\langle\psi(0_+)|\left(S^z\right)^2|\psi(0_+)\rangle_c}$ is the geodesic length of the one-parameter manifold. This already proves the conjecture~\eqref{eq:conjecture}. In Sec.~\ref{subsec:mapping} we formalize and generalize this procedure.

Since the exact analytical expression for $\ell_t$ is rather cumbersome and involves cubic roots, we refrain from showing it here. We can, however, instead check numerically how tight the conjecture bound is. One can evaluate this integral~\eqref{eq:TQSL_3LS} numerically, and e.g.~for $J=1$ and $\lambda_i=-2g=-\lambda_\ast$, we find $T_\mathrm{QSL}g\approx 1.838$, which agrees with the number we obtained using Optimal Control algorithms. Interestingly, this number is smaller than the corresponding one for the 2LS. This means that one can prepare the interacting states faster using a free Hamiltonian. Similarly, one can compute the exact geodesic length $\ell_\lambda'$. Figure~\ref{fig:3LS_nonint} shows the validity of the conjecture at $T=T_\mathrm{QSL}$ as a function of the interaction strength $J/g$.

\section{\label{subsec:mapping}Generalization of the Mapping of Fast-Forward to Counter-Diabatic Protocols}

The previous two examples were very instructive. In particular, we saw that at the quantum speed limit  the protocols which can be obtained by rotating the gauge potential automatically satisfy the conjecture~\eqref{eq:conjecture} because they consist of two pieces: the rotated gauge potential contribution (or more generally rotated counter-diabatic Hamiltonian) and the extra kick contribution due to the rotation, cf.~Eqs.~\eqref{eq:HFF_2LS},~\eqref{eq:HFF_3LS}. The contribution of the first term to $\ell_t$ gives precisely the geodesic length, or more accurately $\int d\lambda \sqrt{g_{\lambda\lambda}}$ along the chosen path $\vec\lambda(t)$, while the second, or kick term results in an extra positive contribution. Hence, for protocols of this form, the conjecture is automatically satisfied. Let us show that this scenario (and hence the validity of the bound) is generic for at least a broad class of fast-forward protocols. To do this, we first prove that any fast-forward protocol can be represented as a rotated counter-diabatic protocol.

Let us assume that there is a Hamiltonian $H_{\rm FF}(t)\equiv H(\lambda(t))$ such that the corresponding wave function $|\psi(t)\rangle$ satisfies the boundary conditions $|\psi(0)\rangle=|\psi_i\rangle$ and $|\psi(T)\rangle=|\psi_\ast\rangle$.  We further assume that $\lambda(t_i)=\lambda_i$ and $\lambda(T)=\lambda_\ast$. The latter assumption is not crucial because $\lambda(t)$ is allowed to change discontinuously (but we assume that $\vert\psi(t)\rangle$ is continuous and differentiable with respect to time). Let us also choose some arbitrary monotonic function $\mu(t)$ which interpolates between $\lambda_i$ and $\lambda_f$ for $t\in [0,T]$ along the adiabatic path, for example:
\[
\mu(t)=\lambda_i+(\lambda_f-\lambda_i) t/T.
\]
We now show that, at the QSL, every fast-forward Hamiltonian can be mapped to a counter-diabatic Hamiltonian.
Define a unitary map $R(t)$ such that
\be
R(t) |\psi(t)\rangle=|\psi_0(\mu(t))\rangle \leftrightarrow |\psi(t)\rangle=R^\dagger(t)|\psi_0(\mu(t))\rangle,
\ee
where $|\psi_0(\mu(t))\rangle$ is the instantaneous wave function and $|\psi(t)\rangle$ is the time-evolved wavefunction under the Hamiltonian $H_\mathrm{FF}(\lambda(t))$.
One can convince oneself that this change-of-frame transformation is not unique but it always exists~\footnote{We note in passing that if we instead have a fast-forward protocol which applies to \emph{all eigenstates} of $H(\mu)$, not just the ground state then we can uniquely (up to phase rotations) define $R(t)$ by applying the relation above to each evolved eigenstate.}. Let us plug the equation above into the Schr\"odinger equation: $i\partial_t |\psi(t)\rangle=H_{\rm FF}(\lambda(t))|\psi(t)\rangle$. Rearranging the terms, we find
\begin{multline}
\left[i (\partial_t R(t)) R^\dagger(t)+ R(t) H_{\rm FF}(\lambda(t)) R^\dagger(t)\right] |\psi_0(\mu(t))\rangle\\
=\partial_t |\psi_0(\mu(t))\rangle.
\nonumber
\end{multline}
Since $\mu(t)$ follows an adiabatic path, the evolution of the instantaneous wavefunction $|\psi_0(\mu(t))\rangle$ \textit{at the QSL} along this path is governed by the gauge potential~\footnote{The gauge potential $\mathcal{A}_\mu$ and the metric tensor $g_{\mu\mu}$ only depend on time implicitly via $\mu(t)$.} 
\begin{equation*}
\partial_\mu |\psi_0(\mu)\rangle =  \mathcal A_\mu |\psi_0(\mu)\rangle.
\end{equation*}
Using these relations, we immediately conclude that 
\begin{multline}
H_{\rm FF}(\lambda(t))=\dot\mu R^\dagger(t) \mathcal A_\mu R(t) -i R^\dagger(t) \partial_t R(t)\\
+ R^\dagger(t) K_\mu R(t),
\label{eq:map_FF_CD}
\end{multline}
where
\[
K_\mu |\psi_0(\mu)\rangle=0.
\]
The last term in Eq.~\eqref{eq:map_FF_CD} does not affect the ground state and reflects the gauge freedom in the choice of the gauge potential we discussed in Sec.~\ref{subsec:3LS_I} above. We can simply absorb it into $\mathcal A_\mu$ via $\mathcal A_\mu\to \mathcal A_\mu+ K_\mu/\dot\mu$. There is also an obvious gauge freedom in choosing the mapping related to the choice of the function $\mu(t)$ and the rotation matrix $R$. In the two simple examples we analyzed above this gauge freedom can be used to make the  second term in Eq.~\eqref{eq:map_FF_CD}:  $-i R^\dagger(t) \partial_t R(t)$ to be orthogonal to the first term, i.e.~to enforce the condition
\be
i\langle \psi_0(\mu(t))| \left\{\mathcal A_\mu, (\partial_t R(t)) R^\dagger(t)\right\}_+ |\psi_0(\mu(t)\rangle=0.
\label{eq:orth_condition}
\ee

Equation~\eqref{eq:map_FF_CD} shows that, at the QSL, any fast-forward Hamiltonian can be written as a rotated counter-diabatic Hamiltonian. Clearly, by applying an appropriate time-dependent phase transformation to $R$: $R(t)\to R(t)\mathrm e^{i \phi(t)}$, with $\phi(t)$ an overall time-dependent scalar phase, we can get a similar mapping of fast-forward to counter-diabatic Hamiltonians also away from the QSL. To do this, in Eq.~\eqref{eq:map_FF_CD} we replace $\dot\mu \mathcal A_\mu\to H_{\rm CD}=H(\mu(t))+\dot\mu \mathcal A_\mu$. 

Once we have establish the equivalence of the fast-forward and counter-diabatic protocols we can examine which conditions we need in order to satisfy the conjecture~\eqref{eq:conjecture}. 
Since
\begin{multline}
\langle \psi(t) |\left[ H_{\rm FF}(\lambda(t))\right]^2 |\psi(t)\rangle_c=\\
\langle \psi_0(\mu(t)) | \left( \dot \mu \mathcal A_\mu-i (\partial_t R(t)) R^\dagger(t) \right)^2|\psi_0(\mu(t) )\rangle_c,
\label{eq:norm_FF_1}
\end{multline}
a sufficient condition for the conjecture is that $R$ can be represented as a finite product of piecewise constant transformations:
\begin{equation*}
R=R_1R_2\cdots R_N
\end{equation*}
where each $R_j$ is time independent and acts only in the interval $[T_{j-1},T_j]$ ($T_N=T$). Then, similar to the 2LS and 3LS examples, within the bulk of each time interval only the rotated counter-diabatic Hamiltonian is effective, while at the interval boundaries the wave function evolves according to the kicks given by $R_j$. This implies that, under the assumption that such representation of $R$ exists, using Eq.~\eqref{eq:norm_FF_1} we have
\begin{multline}
\sqrt{\langle \psi(t) | \left[ H_{\rm FF}(\lambda(t))\right]^22 |\psi(t)\rangle_c}=\\
|\dot\mu | \sqrt{g_{\mu\mu}}
+\sqrt{\langle \psi_0(\mu(t)) | \left(-i (\partial_t R(t)) R^\dagger(t)\right)^2|\psi_0(\mu(t) )\rangle_c}\\
\geq |\dot\mu| \sqrt{g_{\mu\mu}}
\end{multline}
and the conjecture~\eqref{eq:conjecture} follows immediately. At the moment we cannot prove that in general there exists no smooth $R$ such that the integral of expression~\eqref{eq:norm_FF_1} is smaller than the geodesic length. We also do not know any general recipe for finding $R$.

Below, we briefly explain the intuition behind the orthogonality condition~\eqref{eq:orth_condition}.  We again consider a two level system with the Hamiltonian similar to Eq.~\eqref{eq:H_2LS}
\begin{equation}
H(t) = h_0S^z + h_1\cos\phi(t)S^x + h_1\sin\phi(t)S^y,
\end{equation}
but with an important difference that now the control parameter is the azimuthal angle $\phi$. For any $h_0\neq 0$ the parameter space geodesic length $\ell_\phi$ is larger than the distance between wave functions, which is determined by the global geodesic $\mathcal{L}$. Thus, the conjecture gives a tighter bound for $\ell_t$ and hence for the QSL.

Similar to Sec.~\ref{subsec:2LS} we choose the initial and target states lie in the $(x,z)$-plane; they are defined as the ground states at $\phi_i = 0$ and $\phi_\ast = \pi$. Because adiabatic transformations with respect to $\phi$ are generated by rotations around the $z$-axis, the gauge potential in this case is simply $\mathcal A_\phi=S_z$~\cite{kolodrubetz_16}, and hence the counter-diabatic Hamiltonian is
\[
H_\mathrm{CD}=\dot \phi S_z.
\]
It is easy to check that $g_{\phi\phi}=1/2\sin^2(\theta/2)$, where $\tan\theta=h_1/h_0$, leading to the geodesic length \[
\ell_{\phi}=\int_0^\pi \sqrt{g_{\phi\phi}}\, d\phi={\pi\over \sqrt{2}} |\sin(\theta/2)|.
\]
Except for $\theta=\pi/2$ corresponding to $h_z=0$, this length is clearly longer than the geodesic length along the great circle in the $\theta$-direction given by $\mathcal{L} = \ell_\lambda=\theta$ [c.f. Sec.~\ref{subsec:2LS}].

In order to map $H_\mathrm{CD}$ to $H_\mathrm{FF}$ we need to rotate the former around some axis in the $(x,y)$-plane (say the $y$-axis for concreteness) by a time-dependent angle $\gamma(t)$. This rotation defines precisely the operator $R(t)=\exp[-i\gamma(t)S^y]$ from the discussion above, leading to 
\begin{equation}
H'_\mathrm{FF} =\dot \phi\cos\gamma \; S^z + \dot\phi \sin\gamma \; S^x - \dot \gamma S^y.
\label{eq:HFF_2LS_gen}
\end{equation}
As in Sec.~\ref{subsec:2LS}, in order to fix the magnitude of the $h_z$-field we can rescale the time by the factor $\dot\phi \cos\gamma/h_z$ to obtain
\begin{equation}
H_\mathrm{FF} = h_z\; S^z + h_z \tan \gamma \; S^x -h_z {\dot \gamma\over \dot\phi \cos\gamma} S^y.
\end{equation}
Requiring the magnitude of the field transverse to $h_z$  to be fixed at $h_0$ leads to the condition
\[
\tan^2 \gamma+ [\dot\gamma/ (\dot \phi \cos\gamma)]^2=(h_1/h_0)^2,
\]
which can always be satisfied for some $\gamma(t)\in [0,\pi/2]$ as along as we require that $|\dot\gamma|/|\dot\phi| < h_1/h_0 $.

Observe that the last term in Eq.~\eqref{eq:HFF_2LS_gen} [proportional to $S^y$] is always orthogonal to the first two terms [the rotated CD protocol]. Therefore, the orthogonality condition~\eqref{eq:orth_condition} is satisfied for any choice of the protocol $\phi(t)$ and the conjecture $\ell_t>\ell_\phi$ is correct. Note that the analysis above is true for any rotation in the $(x,y)$-plane, not just around the $y$-axis. For this reason the conjecture works for any fast forward protocol.

In the next section, we check the validity of our conjecture numerically using various integrable and non-integrable, local and non-local few- and many-particle systems.

\section{\label{sec:numerical}Numerical Verification of the Geometric Bound Conjecture}

In this section, we use algorithms from Optimal Control to numerically test the geometric bound conjecture in systems where analytical solutions are limited by the complexity arising from the enhanced dimensionality of their Hilbert spaces.

\subsection{\label{subsec:3LS_II}Three-Level System II}

The fast-forward Hamiltonian we found in Sec.~\ref{subsec:3LS_I} is non-interacting. One might wonder how the physics of the 3LS discussed in Sec.~\ref{subsec:3LS_I} changes if we look for an interacting fast-forward Hamiltonian. In other words, as before we start from and target the ground state of $H_\mathrm{3LS}$, see Eq.~\eqref{eq:H_3LS}, for $\lambda_i=-2g=-\lambda_\ast$ and $J=1$, but this time we also evolve with $H_\mathrm{3LS}(t)$. Hence, the fast-forward Hamiltonian for this problem must be in the same control parameter manifold as Eq.~\eqref{eq:H_3LS} for some optimal protocol $\lambda(t)$. Recently, methods from Shortcuts to Adiabaticity were applied to study related setups of three-level systems~\cite{chang_07,sugny_08,chen_12,giannelli_14,li_16,theisen_17}. The physics of this optimization problem below the QSL, i.e.~for $T<T_\mathrm{QSL}$, was analyzed extensively in Ref.~\cite{bukov_17symmbreak}, where it was shown that the state preparation problem close to optimality exhibits genuine quantum control phase transitions as a function the protocol duration $T$, including symmetry breaking, which introduce sharp changes in the functional form of the optimal protocols.

Despite the similarity of the current setup to the one in Sec.~\ref{subsec:3LS_I}, for this initial value problem, we were unable to find the corresponding rotation of the counter-diabatic Hamiltonian to its fast-forward counterpart analytically, cf.~Secs.~\ref{subsec:2LS} and~\ref{subsec:3LS_I}. Nevertheless, the existence of a finite quantum speed limit  can be argued using Optimal Control theorems~\cite{jurdjevic_72} and, variational fast-forward protocols can been constructed which put an upper bound on the QSL~\cite{bukov_17symmbreak}. This motivates the search for an approximate fast-forward protocol $\lambda(t)$ using Optimal Control algorithms. 

Although within the scope of some numerical limitations, Optimal Control allows us to test the validity of the conjecture~\eqref{eq:conjecture}. Indeed, applying GRAPE~\cite{glaser_98,khaneja_01,grape_05}, results in an (almost) optimal protocol which, in turn, defines a proper fast-forward Hamiltonian. To find it, we fix a protocol duration $T$ and discretize time in $N_T=100$ equal steps. We then use GRAPE, which is based on gradient ascend, to find the best possible value for the control field $\lambda(t)$ at each time step in the range $\lambda(t)\in[-16g,16g]$,which optimizers the fidelity of being in the target state at the end of the protocol $t=T$. In order to minimize the probability of getting stuck in a local fidelity maximum, we repeat the procedure a total of two hundred times and post-select the best outcome. 

\begin{figure}[t!]
	\includegraphics[width=1.05\columnwidth]{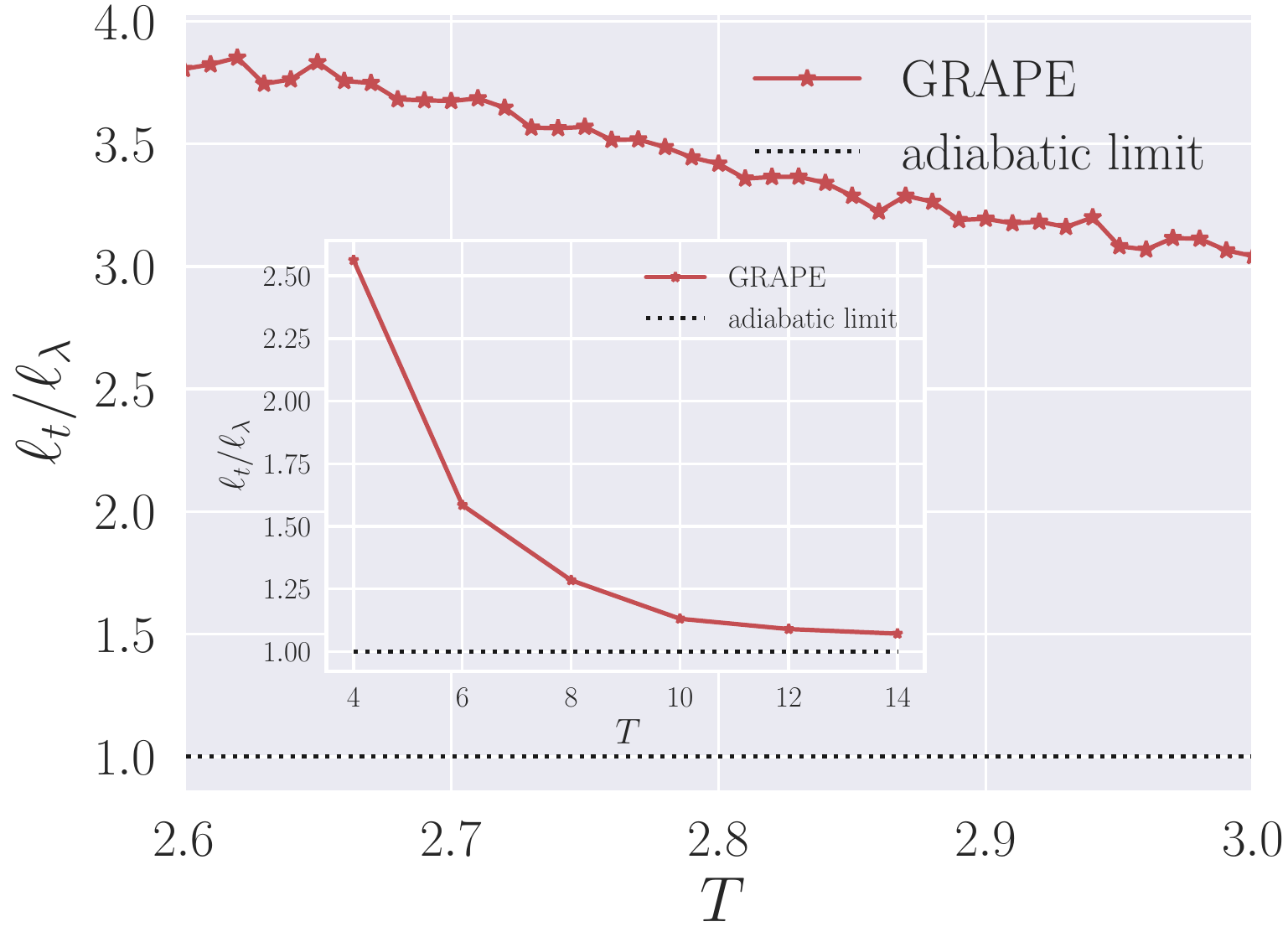}
	\caption{\label{fig:3LS_int} Numerical justification of the conjecture, cf.~Eq.~\eqref{eq:conjecture}, across the quantum speed limit of preparing the interacting ground state of the Hamiltonian $H_\mathrm{3LS}$ following evolution generated by $H_\mathrm{3LS}(t)$. The inset shows numerical evidence that the conjecture is saturated in the adaibatic limit. The parameters are $\lambda_i/g=-2=-\lambda_\ast/g$. The optimal control algorithm used is GRAPE~\cite{glaser_98,grape_05}.}
\end{figure}

The optimal protocol enables us to test the geometric bound conjecture~\eqref{eq:conjecture} numerically. To this end, we first identify the quantum speed limit  within numerical precision, which allows us to safely focus on protocol durations $T>T_\mathrm{QSL}$ [note that the conjecture holds only above the QSL, where we can achieve unit fidelity]. In this regime, we also make sure that the approximate fast-forward protocol indeed prepares the target state with fidelity $F_h(T)=\vert\langle\psi(T)|\psi_\ast\rangle\vert^2$ of at least $99.99\%$. To evaluate the LHS $\ell_t$, we use the fast-forward Hamiltonian with $\lambda(t)$ obtained using GRAPE. The quantity $\ell_t$, related to the time-averaged of the square root of the energy fluctuations of $H_\mathrm{FF}$, is then computed numerically. On the RHS of~\eqref{eq:conjecture}, we determined $\ell_\lambda$ independently by using (i) the geodesic length computed from the analytical gauge potential~\eqref{eq:A_3LS} and (ii) -- a very slow ramp in the adiabatic limit ($T=100J$), where the bound is saturated. We found excellent agreement between the two approaches. Figure~\ref{fig:3LS_int} shows the result which confirms the validity of the geometric bound conjecture for the interacting 3LS setup. 

The inset to Fig.~\ref{fig:3LS_int} shows numerical evidence that the inequality is saturated in the adiabatic limit. Checking this is a nontrivial task, because at late times there are many protocols with unit fidelity.  Most of these optimal protocols have large energy fluctuations and will not be close to saturating the bound. To circumvent this issue, we initiated GRAPE with a smooth adiabatic protocol. 
The unit-fidelity protocols we obtained this way smoothly connect to the adiabatic solution in the limit $T\to\infty$, where we proved that they saturate the bound.
Interestingly, any smooth deformations on top of the adaibatic protocol introduced by GRAPE, always lead to $\ell_t/\ell_\lambda>1$, providing additional evidence that the conjecture is valid. The same applies to more complex many-body systems, see also Fig.~\ref{fig:FullIsing_length} and Fig~\ref{fig:SYK}.

\subsection{\label{subsec:Ising}Nonintegrable Ising Chain: Ground State Physics}

The previous examples we discussed all share in common a few-dimensional Hilbert space. A natural question to ask is whether the Conjecture~\eqref{eq:conjecture} holds for many-body systems. In this section, we study a non-integrable Ising chain with emphasis on the ground state physics. Nonintegrability here implies both the absence of a closed-form solution for the gauge potential, and the presence of locally thermalizing quantum dynamics which obeys the Eigenstate Thermalization Hypothesis (ETH)~\cite{ETH_review}. Hence, this model represents a generic quantum many-body system, and our goal below is to test the geometric bound conjecture~\eqref{eq:conjecture} on it.  

Consider the non-integrable transverse-field Ising model (TFIM) in a longitudinal field, described by the Hamiltonian 
\begin{equation}
H(t) = -\sum_j JS^z_{j+1}S^z_j + g S^z_j + \lambda(t) S^x_j.
\label{eq:H_MB}
\end{equation}
In the following, we set $J=1$ as a reference energy scale. Once again, $\lambda(t)$ denotes the control field. The initial and target states are the interacting ground state for $\lambda_i=-2g=-\lambda_\ast$, respectively, and the protocol duration is denoted by $T$. Quantum state preparation in this setup has been studied extensively using Reinforcement Learning in Ref.~\cite{bukov_17RL}, and this state preparation problem has been shown to have glassy optimization complexity~\cite{day_17}. Due to the lack of a closed-form solution of the stationary Schr\"odinger equation, it is not possible to obtain a the ground state manifold of the system as a function of $\lambda$ analytically. Therefore, we restrict the analysis of this initial value problem to the methods of Optimal Control.

Because of the extensivity of the spectrum of many-body systems, it is unphysical to allow for unbounded drive fields $\lambda(t)$, since local control does not grant access over extensively large energy scales. Therefore, we consider the experimentally relevant situation of a bounded drive $\lambda(t)\in[-4g,4g]$. As before, we discretize the protocol duration $T$ in time steps $\delta t$, and study the problem using two different control algorithms (see Sec.~\ref{subsec:3LS_II} for details):
(i) GRAPE looks for continuous protocols, while (ii) Stochastic Descent (SD) has proven useful to look for the so-called bang-bang protocols, i.e.~protocols which take values on boundary of the allowed domain: $\lambda\in\{\pm4\}$. Although discontinuous, the family of bang-bang protocols are known to contain an optimal solution as a consequence of Pontryagin's maximum principle.

It is not known what the quantum speed limit  for this problem is, nor whether it is finite in the thermodynamic limit. Therefore, we make sure to consider only optimal protocols with durations $T$, which allow for enough time to prepare the target state with \emph{many-body} fidelity $F_h(T)=\vert\langle\psi(T)|\psi_\ast\rangle\vert^2$ of at least $99.99\%$. In this respect, it is important to mention that close to optimality finite-size effects have been shown to be negligible for this problem setup, starting from a system size of $L>6$ sites, see Ref.~\cite{bukov_17RL}, and hence we restrict to $L=10$ for the results presented here. 

To check the geometric bound conjecture~\eqref{eq:conjecture}, we compute numerically the LHS and RHS. Once the optimal fast-forward protocol $\lambda(t)$ has been determined, the numerical computation of $\ell_t$ on the LHS is straightforward. On the RHS, we can no longer calculate the geodesic length exactly, since we do not have the exact expression for the adiabatic gauge potential. Nevertheless, as we argued in Sec.~\ref{sec:conjecture} and verified numerically in Sec.~\ref{subsec:3LS_II}, we can obtain the geodesic length $\ell_\lambda$ from evolution in the adiabatic limit. 

\begin{figure}[t!]
	\includegraphics[width=1.05\columnwidth]{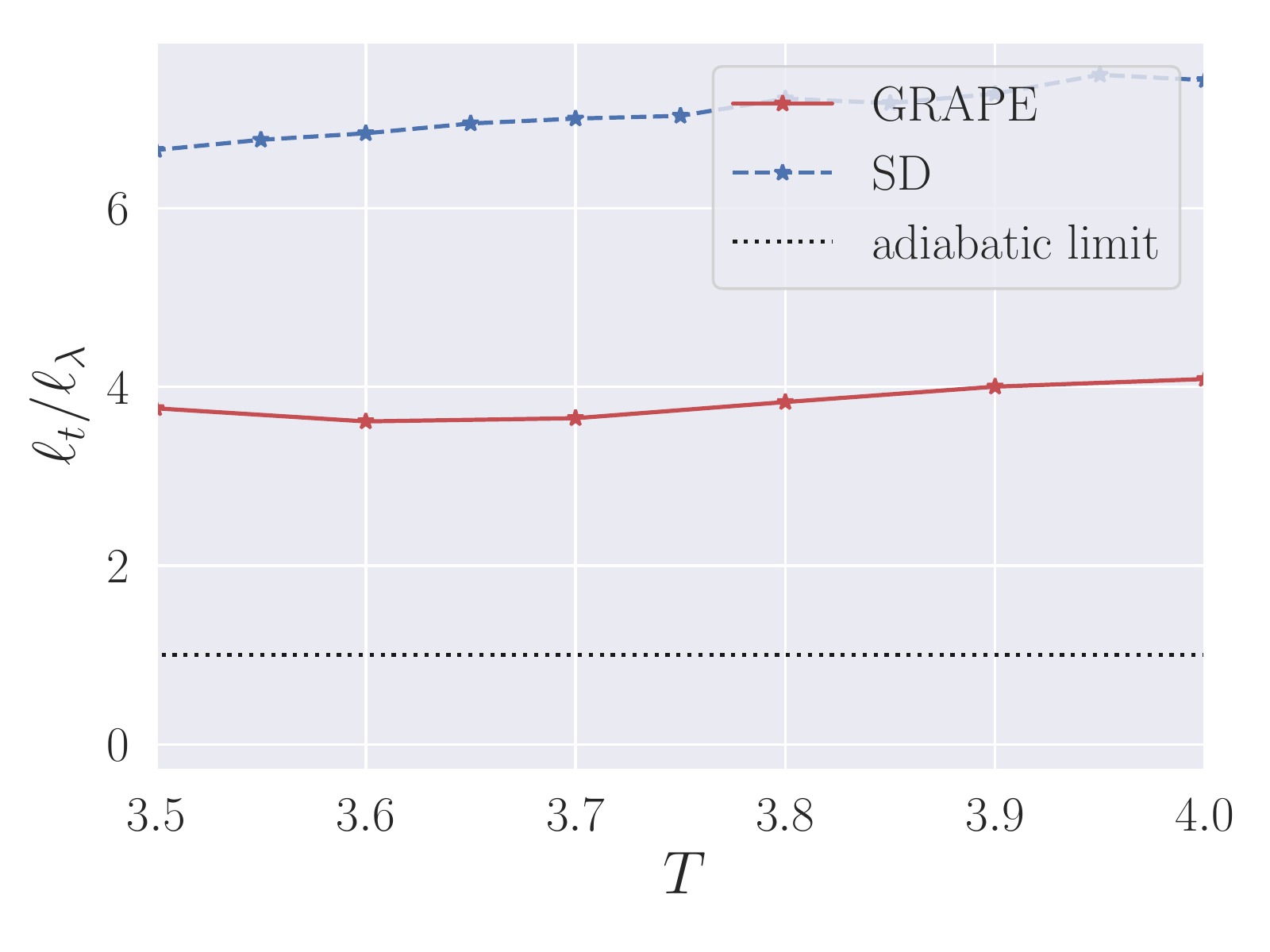}
	\caption{\label{fig:many-body} Numerical justification of the conjecture, cf.~Eq.~\eqref{eq:conjecture},  in the high-fidelity region of the quantum state preparation problem in the many-body Hamiltonian~\eqref{eq:H_MB} as a function of the protocol duration $T$. The parameters are $\lambda_i/g=-2=-\lambda_\ast/g$, $J/g=1$ and $L=10$. We used Stochastic Descent (SD) to find (nearly) optimal bang-bang protocols $\lambda(t)\in\{\pm 4\}$ of time step is $\delta t=0.005J$.}
\end{figure}

Figure~\ref{fig:many-body} shows the ratio $\ell_t/\ell_\lambda$ between the time-integral over the square root of the energy fluctuations of the fast-forward Hamiltonian corresponding to the optimal protocol and the geodesic length, as a function of the protocol duration $T$. It is an interesting observation that, even though both the bang-bang protocols (dashed line) and the continuous GRAPE protocols (solid line) satisfy the conjecture, the average energy variance $\ell_t$ is kept smaller by the GRAPE protocols. We recall that, according to Pontryagin's maximum principle, one can find a bang-bang protocol to achieve (at least) the same fidelity as with any continuous protocol. We attribute the fact that the two families of protocols differ in terms of the average energy variance they create during the evolution, to their robustness properties: while bang-bang protocols might be optimal they have recently been shown to be unstable to small perturbations~\cite{bukov_17RL}. Mathematically bang-bang protocol result in a larger energy variance and hence larger $\ell_t$ because they the Hamiltonian changes very rapidly between the bangs, while the state does not have time to follow. If we associate $\ell_t$ with the fluctuating energy cost following Ref.~\cite{funo_17} then clearly bang-bang protocols are more costly than smooth protocols. Notice that our numerical results suggest that away from the adiabatic limit the conjecture is not tight and the ratio $\ell_t/\ell_\lambda>1$, though in most cases it remains close to one. If we increase protocol times then as expected the ratio $\ell_t/\ell_\lambda$ approaches unity, see e.g. Fig.~\ref{fig:FullIsing_spectrum}.

\subsection{\label{subsec:excited_states}Nonintegrable Ising Chain: Excited States Physics}

It is well known that some properties of low-energy states differ significantly from those of their excited states counterparts. Most notably, in many systems, the ground state physics is protected by a finite gap in the energy spectrum, which renders the adiabatic limit well-defined. In contrast, the energy level spacing for excited states is usually exponentially suppressed in the system size, and for spin-$1/2$ chains scales as $2^{-L}$. Consequently, the time scales for the adiabatic limit are exponentially longer for excited states. On the other hand, fast-forward protocols are allowed to excite the system during the evolution before they prepare the target state. One can imagine harnessing this additional freedom to improve on the time scales for adiabatic state preparation. This raises the question whether the conjecture~\eqref{eq:conjecture} is not violated for excited states. 

\begin{figure}[t!]
	\centering
	\includegraphics[width=1\linewidth]{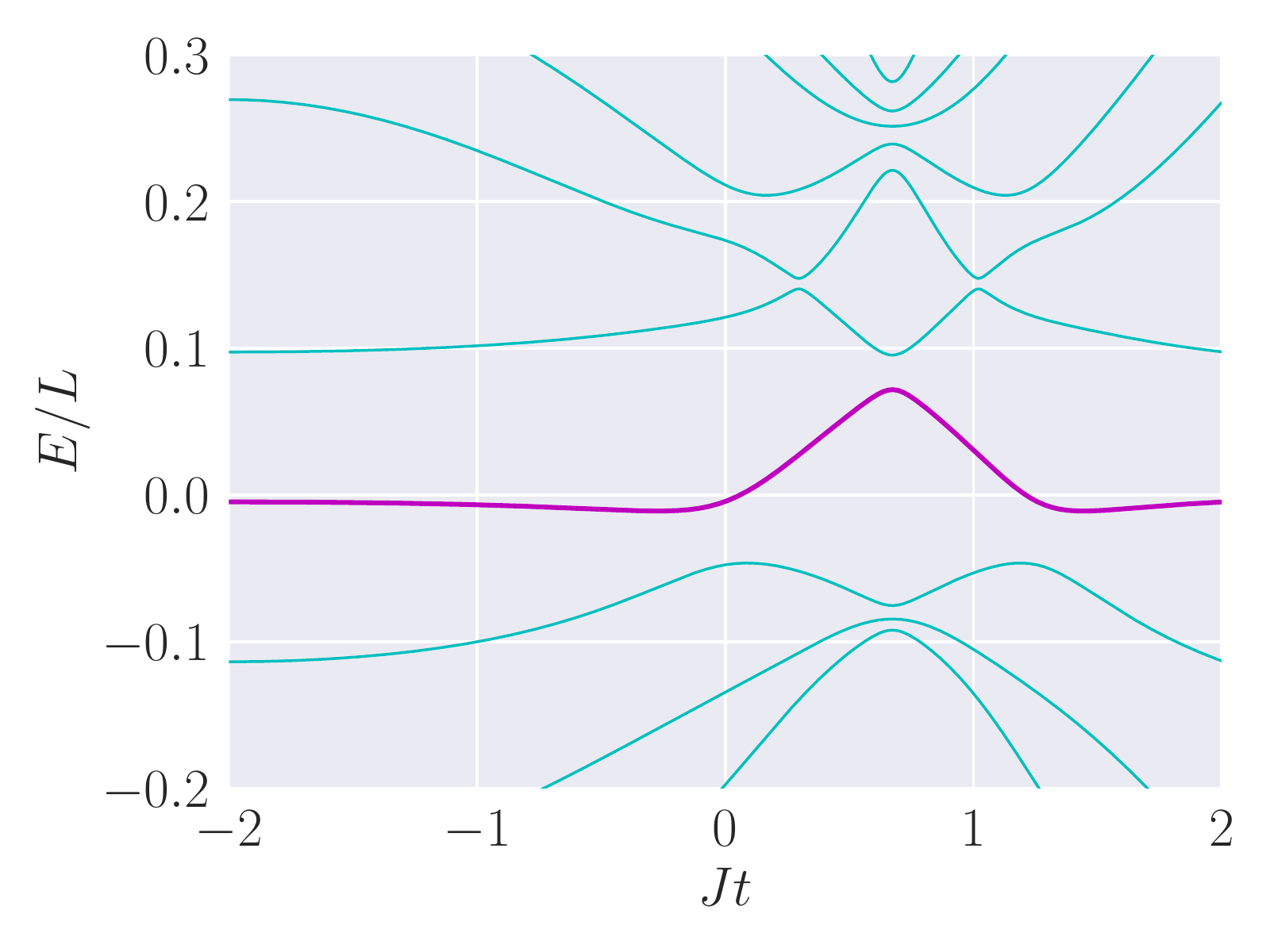}
	\includegraphics[width=1\linewidth]{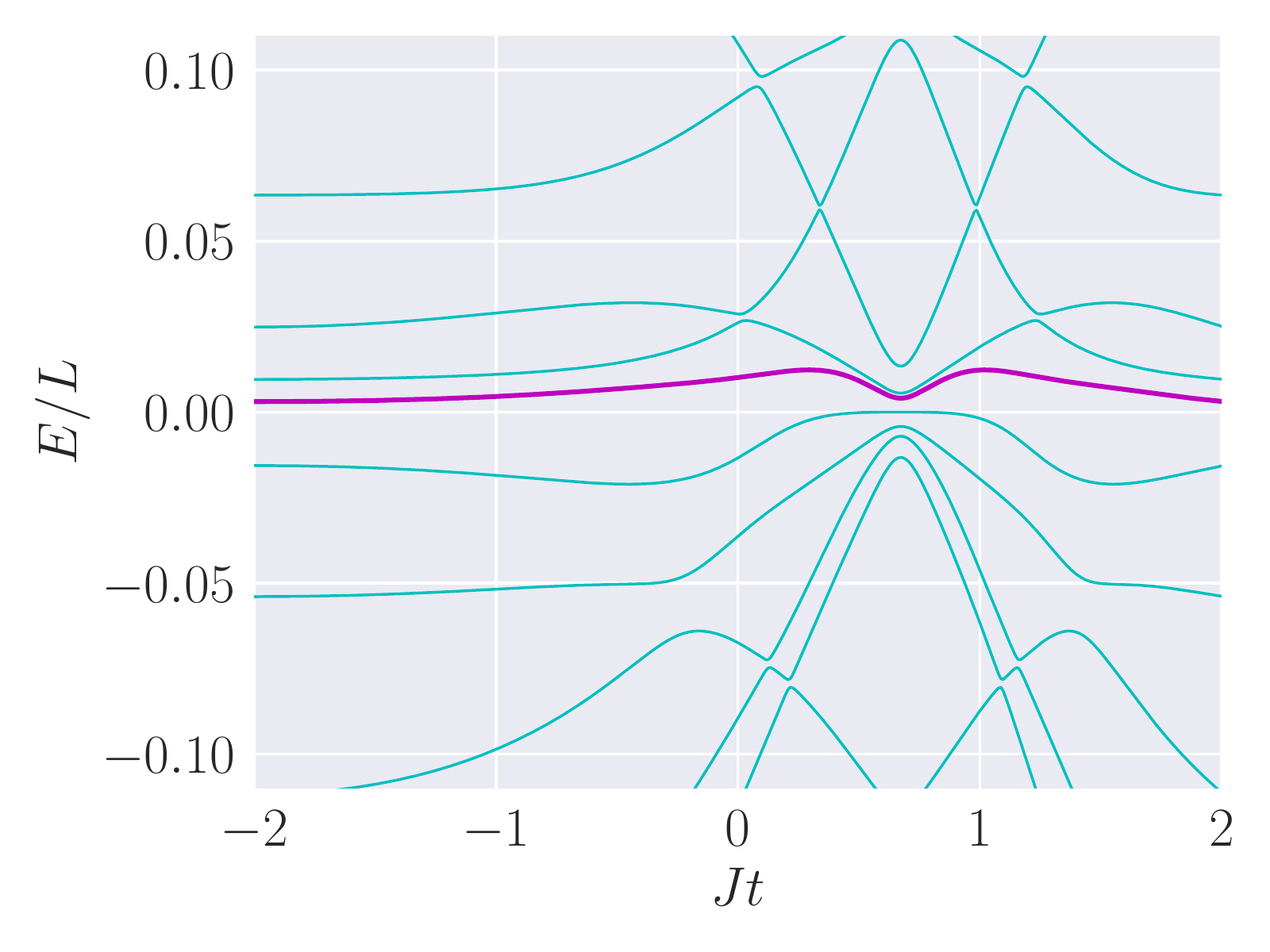}
	\caption{\label{fig:ES_spectra}Excited states of the Hamiltonian~\eqref{eq:H_MB-y_field} along the adiabatic trajectory $\lambda(t)=2\lambda_\ast\cos(\pi t/(2T)) - \lambda_i$ in the vicinity of the adiabatically connected state (magenta) for $L=6$ (up) and $L=8$ (down).}
\end{figure}

To test this, we add a small $y$-field to the non-integrable Ising chain and consider the Hamiltonian:
\begin{equation}
H(t) = -\sum_j JS^z_{j+1}S^z_j + g S^z_j + \lambda(t) S^x_j + h S^y_j,
\label{eq:H_MB-y_field}
\end{equation}
with $h/J=-0.1$, $g/J=1$, and a driving protocol $\lambda(t)/J\in[-2,2]$. We pick for an initial state an infinite-temperature state, characterized by energy which is closest to zero at $\lambda_i=-2J$, see Fig.~\ref{fig:ES_spectra} (purple line). The target state is the adiabatically connected state at $\lambda_\ast=2J$. This choice for the initial state is motivated by ETH, according to which the states in the middle of the spectrum are the first one which become chaotic and lead to thermalization of the system under generic dynamics like dynamics governed by the Hamiltonian~\eqref{eq:H_MB-y_field}. To ensure a finite geodesic length $\ell_\lambda$ and a well defined adiabatic limit, we introduced a small magnetic field in the $y$-direction which breaks the emergent integrability of the system (see clustering of the states closer to $\lambda=0$, i.e.~$Jt\approx 0.8$), and opens up the corresponding unavoided crossings along the adiabatic trajectory.

To test the conjecture for excited states, we consider two spin chains of length $L=6$ and $L=8$, respectively. Imposing periodic boundary conditions, the only two symmetries in the Hamiltonian~\eqref{eq:H_MB-y_field} are translation invariance and parity (reflection about the middle of the chain). Without loss of generality, we work in the zero-momentum sector of positive parity, containing the GS, which allows us to consider only those states that are coupled during the time evolution. The corresponding symmetry-reduced Hilbert sub-spaces have sizes $\mathrm{dim}\mathcal{H}=13$ and $\mathrm{dim}\mathcal{H}=30$, respectively.

\begin{figure}[t!]
	\centering
		\includegraphics[width=1\linewidth]{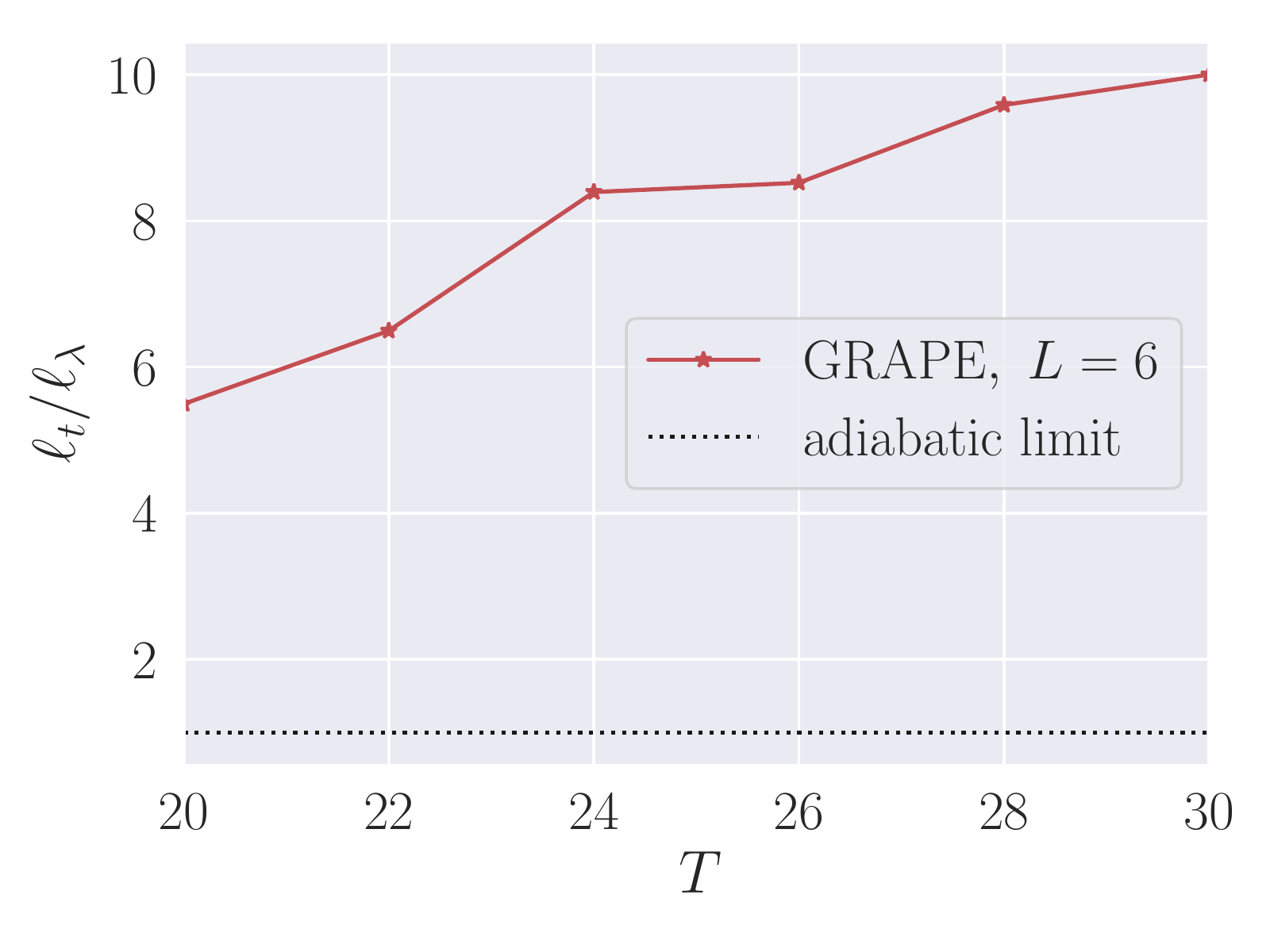}
		\includegraphics[width=1\linewidth]{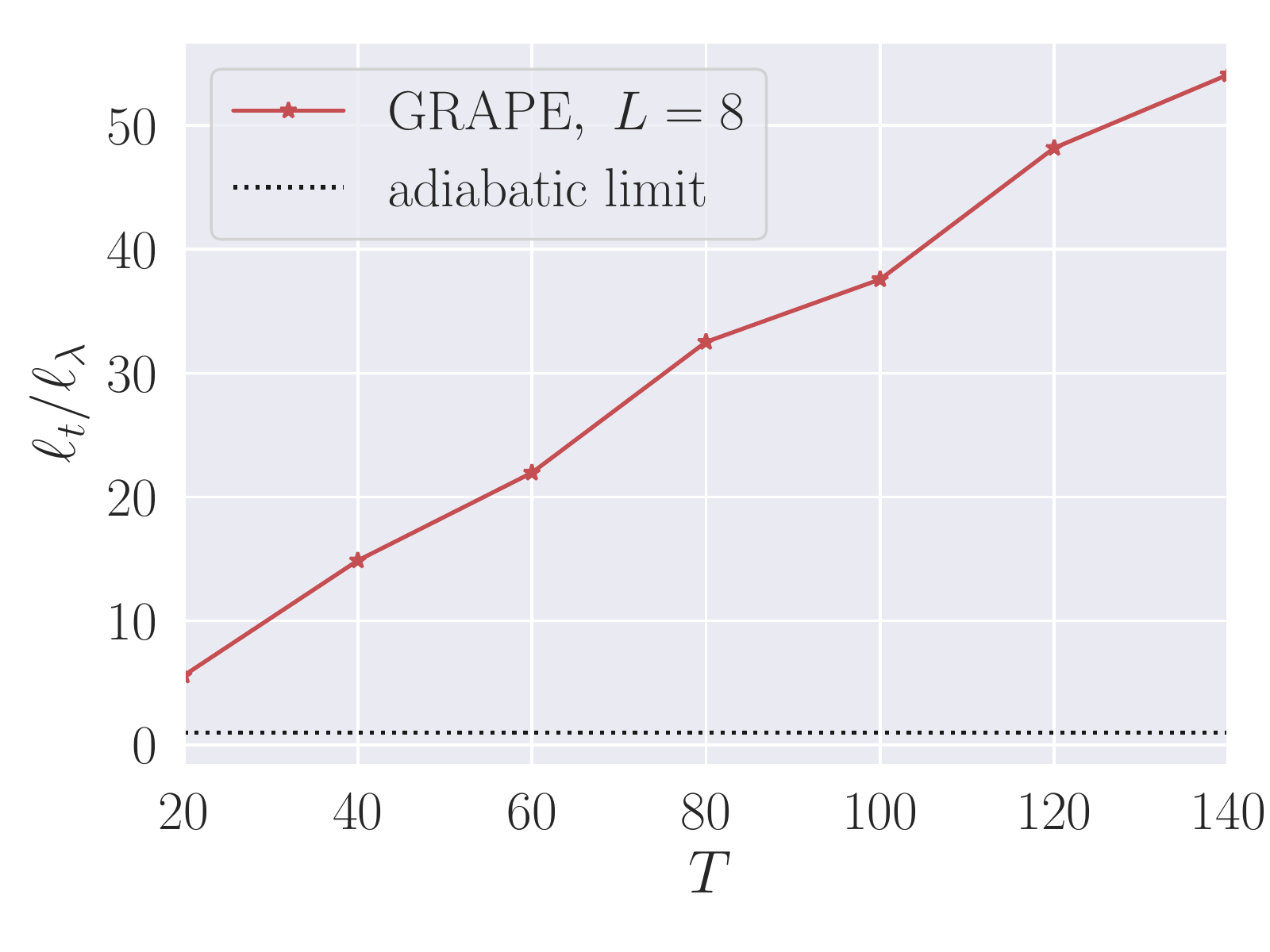}
	\caption{\label{fig:ES_conjecture}Numerical verification of the geometric bound conjecture~\eqref{eq:conjecture} for the excited states of the Hamiltonian~\eqref{eq:H_MB-y_field}.}
\end{figure}

Figure~\ref{fig:ES_spectra} shows parts of the instantaneous energy spectrum of the model, including the adiabatic trajectory from the initial into the target state. The magenta line in the middle marks the adiabatically connected state. One can clearly observe a number of avoided crossings, which are responsible for large protocol durations required to find te system in the adiabatic limit. For instance, to prepare the target state with $99.999\%$ probability adiabatically in the Hamiltonian~\eqref{eq:H_MB-y_field} requires ramp durations on the order of $T=4\times 10^4$ for $L=6$, and $T=10^5$ for $L=8$.

To compute the LHS of the geometric bound conjecture~\eqref{eq:conjecture}, we used GRAPE to find an (almost) optimal protocol sequence of $100$ time steps at a number of fixed protocol durations of order $JT\sim \mathcal{O}(10)$. The non-adiabatic character of these protocols allows for a protocol duration much shorter than the adiabatic ones, yet we made sure that all GRAPE protocols prepare the target state with at least $99\%$ fidelity. 

Figure~\ref{fig:ES_conjecture} demonstrates that the Conjecture~\eqref{eq:conjecture} holds even for the excited states of generic many-body models. Interestingly unlike in the two-spin case the ratio $\ell_t/\ell_\lambda$ increases with the protocol time $T$, c.f. Fig.~\ref{fig:3LS_int}. As we argued we anticipate that in the limit $T\to\infty$ the bound is saturated for any state, ground or excited, because of applicability of APT, so the ratio $\ell_t/\ell_\lambda$ should go down with $T$. The growth of $\ell_t/\ell_\lambda$ is attributed to the numerical GRAPE procedure. Since the energy fluctuations are not part of the cost function, GRAPE is ignorant to them.  as we initiate the algorithm from a random protocol configuration, it flows to a nearby local minimum in the control landscape, which is more likely to have large energy fluctuations with increasing time. Therefore, if we want to use GRAPE to study the adiabatic limit, one should either start close to it, or bias the algorithm towards it.

Nevertheless we clearly see that the inequality $\ell_t>\ell_\lambda$  holds at all  protocol times considered. This result comes with an important consequence. In generic systems satisfying ETH, the geodesic length $\ell_\lambda$ for excited states exponentially diverges with the system size $L$ so the conjecture implies that {\em any} fast-forward protocol is exponentially long.

Since fast-forward protocols excite the system in the basis of the instantaneous Hamiltonian [before they de-excite it to reach the target state with unit probability], one may na\"ively think that by using such out-of-equilibrium protocols it is possible to circumvent the restrictions in the adiabatic limit imposed by the size of the energy gaps in the vicinity of the adiabatically-connected state. However, the validity of the geometric bound conjecture shows that this is not the case. Hence, equilibrium properties impose geometric constraints on the out-of-equilibrium dynamics.

\subsection{\label{subsec:fully_connected}Fully-Connected Ising Model}

\begin{figure}[t!]
	\centering
		\includegraphics[width=1\linewidth]{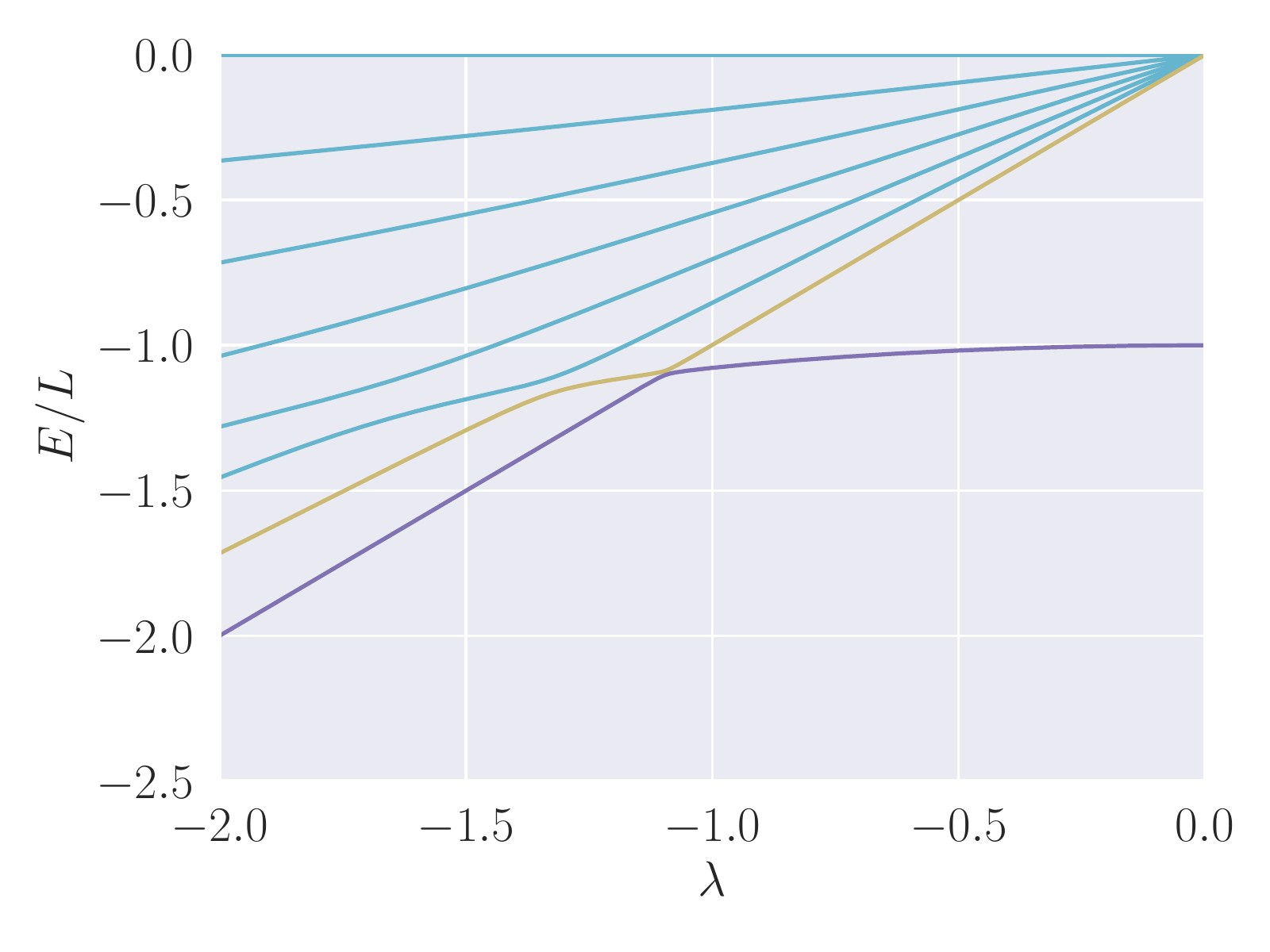}
		\includegraphics[width=1\linewidth]{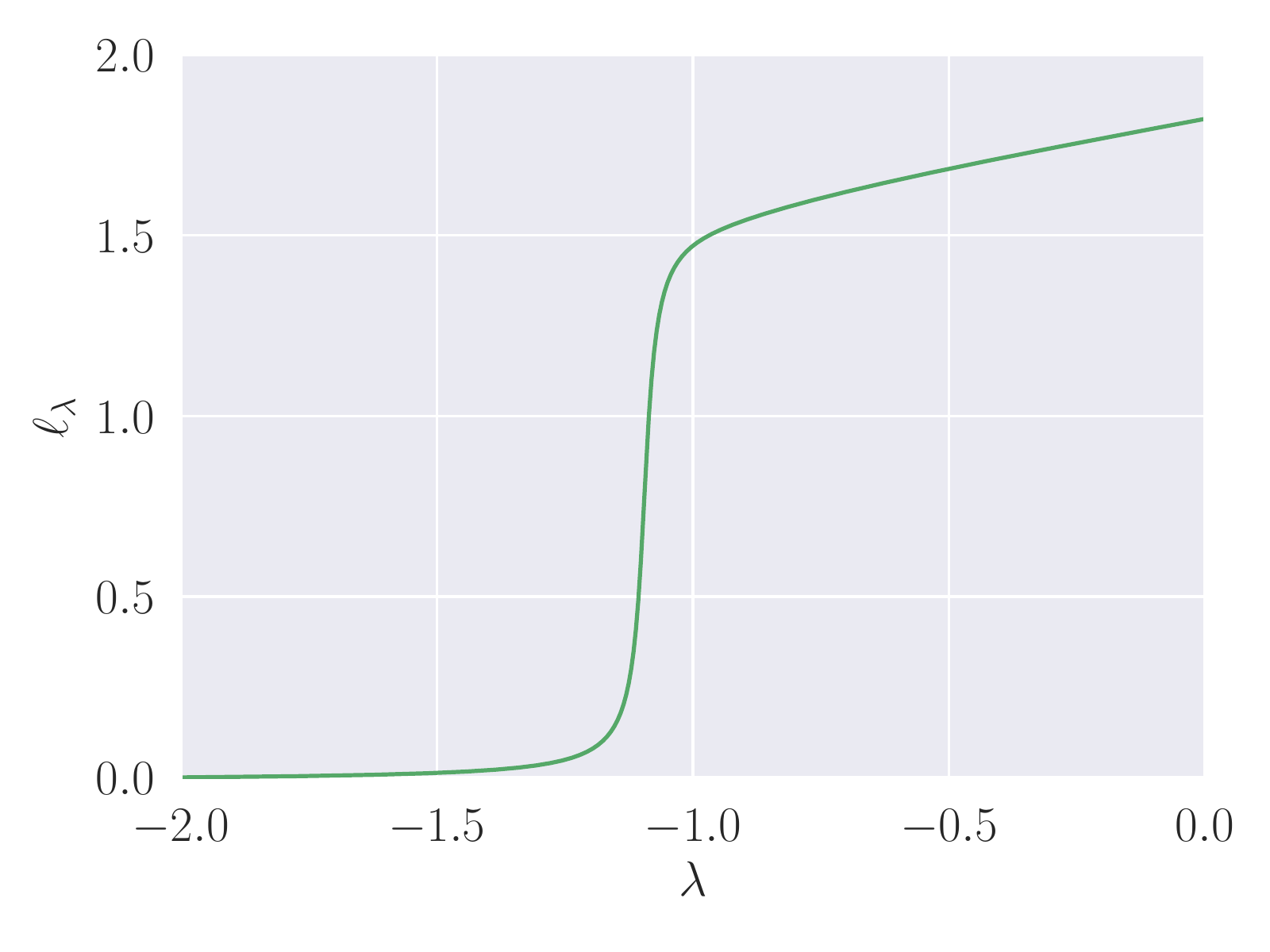}
	\caption{\label{fig:FullIsing_spectrum} Low energy spectrum of a 14-spin disorder free $p$-spin model for $p=51$ discribed by Hamiltonian \eqref{eq: Hpspin}. The model has a first order transition with an exponentially small gap separating the two phases. In the thermodynamic limit, the geodesic length $\ell_\lambda$ jumps by $\pi/2$ at the critical point.  }
\end{figure}

Potential candidates that violate the conjecture are Hamiltonians which have small ground state gaps along their adiabatic path but have a lot of symmetry such that the ground state phases are trivially found by inspection. In those cases one could wonder whether numerical methods from optimal control theory can find protocols that violate our conjecture. Here we check one example and show that it does not. Consider a quantum $p$-spin model without disorder:
\begin{equation}
H=- \frac{L}{2}\left ( \frac{2}{L} \sum_{i=1}^L S^z_i \right)^p+\lambda \sum_{i=1}^L S^x_i, 
\label{eq: Hpspin}
\end{equation}
For any $p>2$, this model has a mean-field like first order transition from paramagnet to ferromagnet with a gap~\cite{Jarg2010} exponentially closing with the system size $L$ This makes it hard to adiabatically cross the transition but at the same time the ground states in the two phases are trivial $Z$ and $X$ polarized product states. Note that the ground state is unique for any odd $p$. Moreover, the Hamiltonian conserves total angular momentum $S^2$ such that the effective Hilbert space dimension is only $L+1$. Figure~\ref{fig:FullIsing_spectrum} shows the low energy spectrum of an $L=14$ spin model for large $p$.  Even though the gap closes exponentially, the geodesic length does not exponentially grow with system size. In contrast, in the thermodynamic limit, it undergoes a jump of $\pi/2$ at the critical point. The latter reveals the simple Landau-Zener nature of the problem, with essentially only two states participating in the transition. 

\begin{figure}
	\includegraphics[width=\columnwidth]{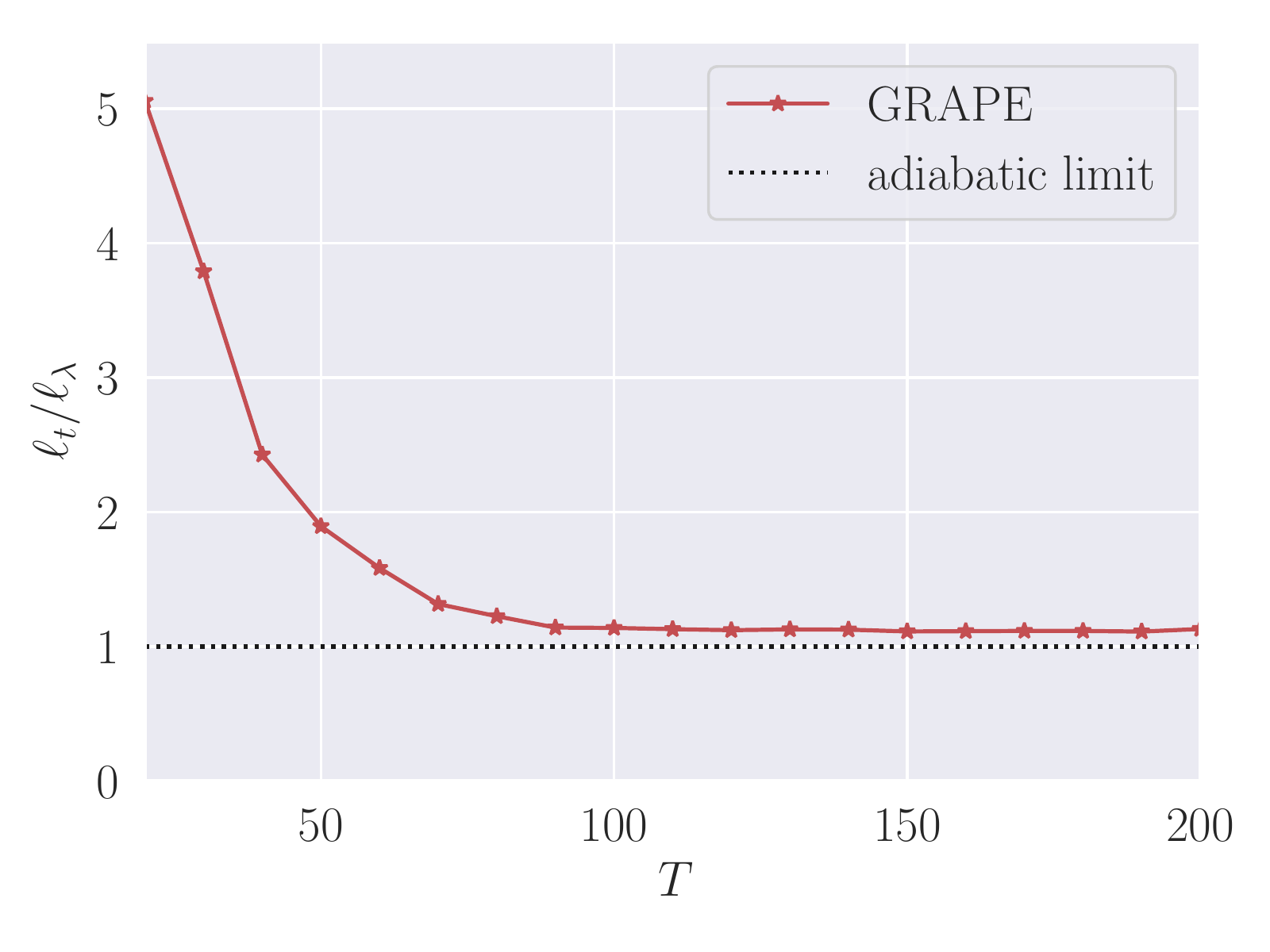}
	\caption{\label{fig:FullIsing_length} Numerical verification of the geometric bound conjecture~\eqref{eq:conjecture} for the ground state of a disorder free $p$-spin model described by Hamiltonian~\eqref{eq: Hpspin}.}
\end{figure}

Once again we use GRAPE to numerically find close-to-unit fidelity protocols that cross the quantum phase transition, i.e.~they start at $\lambda=-2$ and end at $\lambda=0$. The small gap and the highly non-local nature of the Hamiltonian seem to make the optimal control problem significantly harder than any other models considered so far in this work. Typical protocols, obtained from a random initial seed for the GRAPE, have energy fluctuations which are about two orders of magnitude larger than the conjecture bound~\eqref{eq:conjecture}. In order to obtained good protocols with small energy variance we therefore bias the algorithm to the right corner of phase space by starting from the geodesic protocol with some small random noise part. This results in much better protocols which, in the adiabatic limit saturate the bound, see Fig.~\ref{fig:FullIsing_length}. For shorter times, when the inverse time becomes comparable to the minimum gap along the trajectory, we can still find almost-unit-fidelity protocols but their energy variance grows rapidly with decreasing time. Numerical optimal control results thus suggest that our conjecture is also satisfied for mean-field like first order quantum phase transitions.

\subsection{\label{subsec:SYK}Free fermions to SYK model}

So far, all examples were disordered free and, apart from the fully connected model in section~\ref{subsec:fully_connected}, they were also local. While this covers most most of the physically realizable Hamiltonian in experiments, there are some interesting non-local models with quenched disorder.

It is important to check the validity of the conjecture in a non-local setup. Let us therefore consider a Hamiltonian which interpolates between Sachdev-Ye-Kitaev (SYK) model and free fermions,
\begin{equation}
H(t)=\lambda(t)\sum_{j=1}^L \left(c^\dagger_{j+1}c_j +h.c. \right) +\sum_{i,j,k,l=1}^L U_{ijkl} c^\dagger_i c^\dagger_j c_k c_l,
\label{eq:HSYK}
\end{equation}
where $U_{ijkl}$ is a random variable drawn from a normal distribution with zero mean and variance $L^{-3/2}$, and $\lambda(t)$ is the drive. Here $L$ labels the number of fictitious sites in the fully-connected quantum dot, and $c^\dagger_j$ creates a spinless fermion on such a site $j$. The free-particle kinetic energy (hopping) term is assumed to have periodic boundary conditions.

\begin{figure}
	\includegraphics[width=\columnwidth]{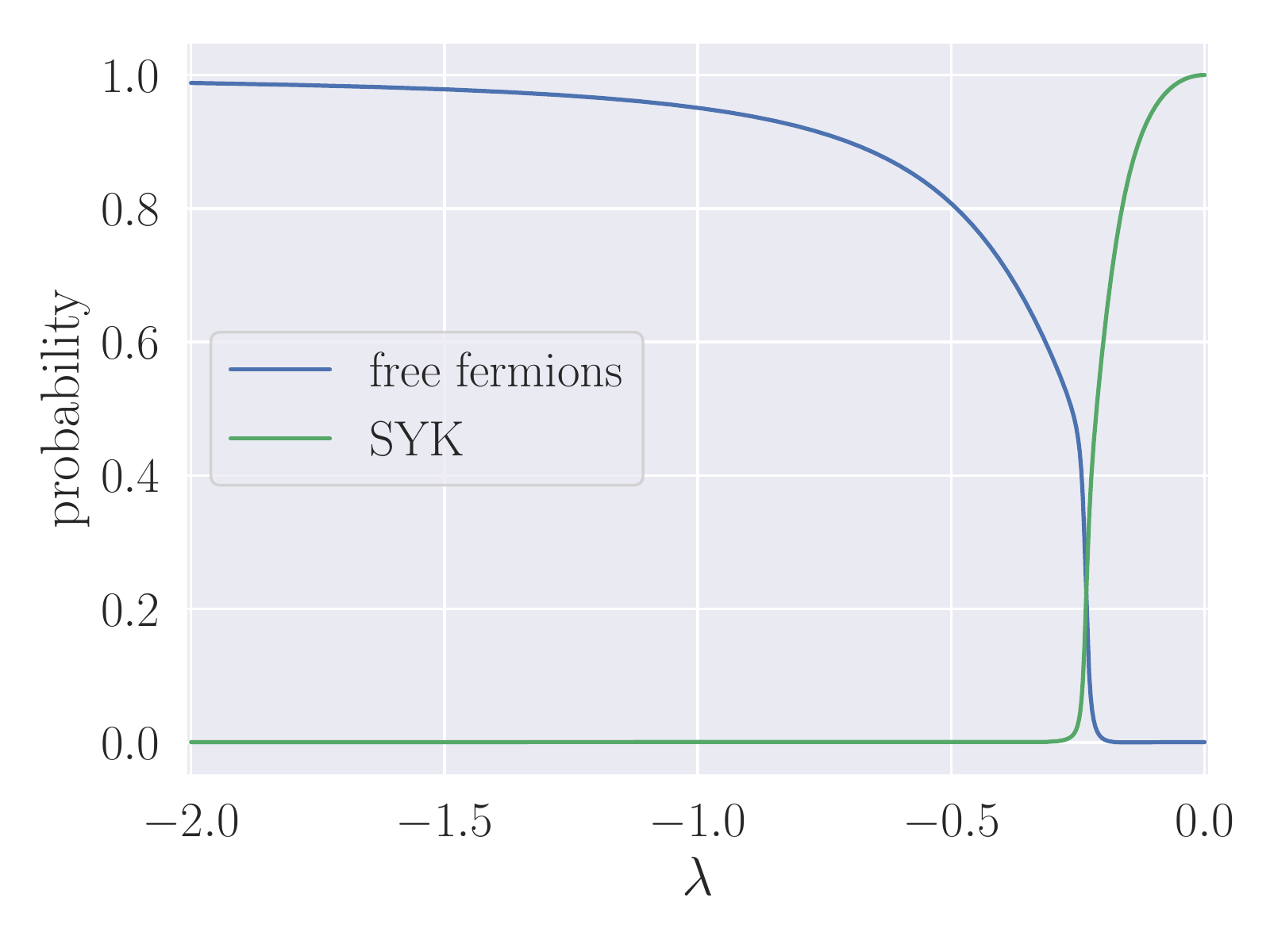}
	\caption{\label{fig:SYK_overlap} Probability to find the ground state of Hamiltonian~\eqref{eq:HSYK} at a particular value of $\lambda$, in the free fermion state $\lambda=\infty$ (blue line) and the SYK ground state $\lambda=0$ (green line). The data shows a single typical realization of the disorder for half filling at $L=8$.}
\end{figure}

For $\lambda=\infty$, the resulting local noninteracting model is described by free fermions, while for $\lambda=0$ it becomes the SYK model. Numerically we initialize the system in the ground state of $\lambda=-2$, which has high overlap with the non-interacting ground state and can thus be considered in the Fermi-liquid phase, see Fig.~\ref{fig:SYK_overlap}. We target the SYK ground state at $\lambda=0$ for a single realization of the disorder. For a given disorder realization there is a sharp transition from a Fermi liquid to a non-Fermi liquid at a critical value of the hopping. Like in the fully-connected Ising model, this is accompanied with a sharp jump in the geodesic length.

\begin{figure}
	\includegraphics[width=\columnwidth]{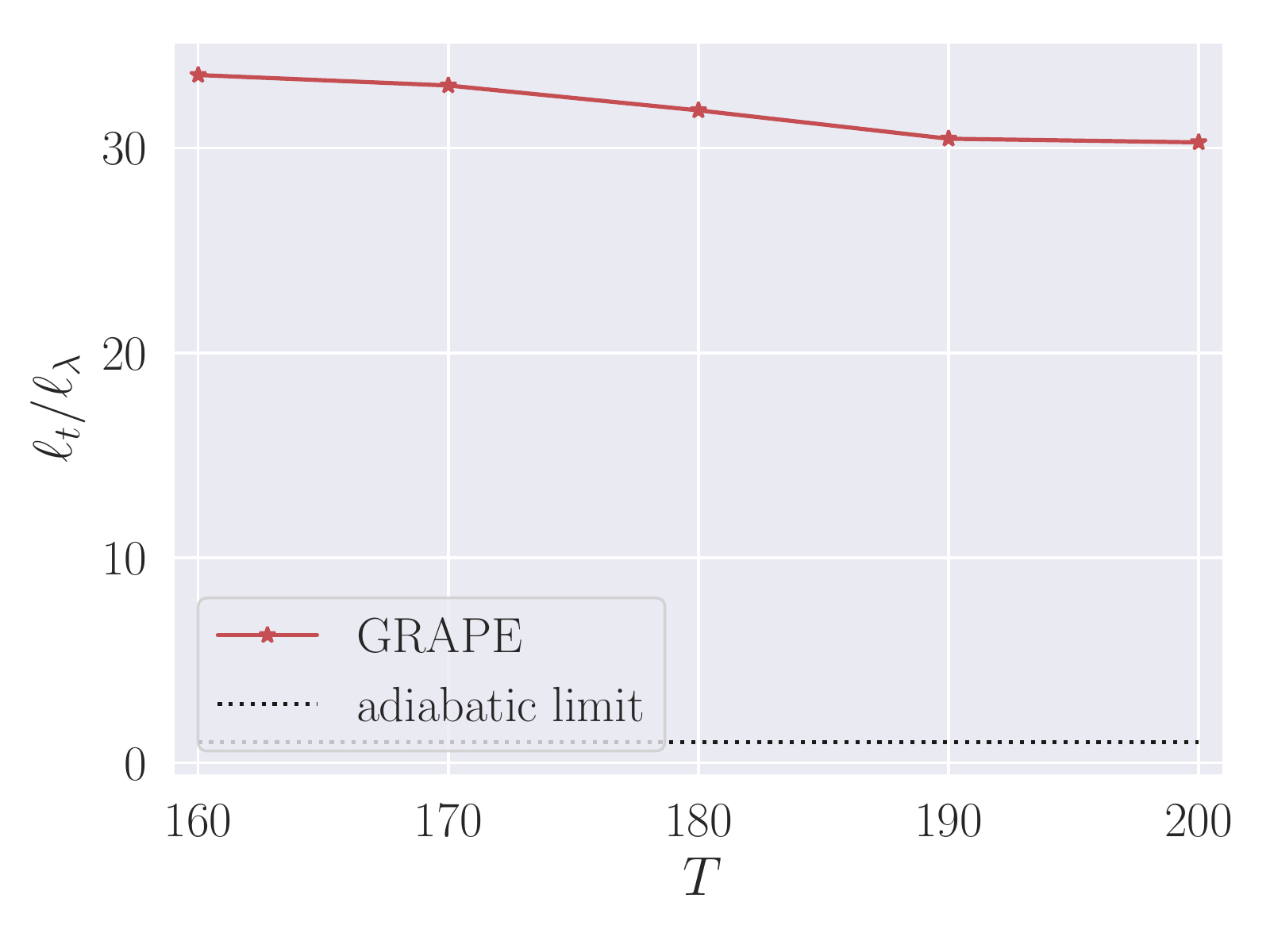}
	\caption{\label{fig:SYK} Numerical verification of the geometric bound conjecture~\eqref{eq:conjecture} for the ground state of a fermionic SYK model described by Hamiltonian~\eqref{eq:HSYK}. The data shows a single disorder realization, since every realizations comes with its own optimal protocols. For the presented realization the quantum speed limit is estimated to be around $T=150$. The inset shows numerical evidence that the conjecture is saturated in the adaibatic limit. The data shows a single typical realization of the disorder for half filling at $L=8$.}
\end{figure}

As before, we use GRAPE to find nearly optimal fast-forward protocols of duration $T$, and verify numerically the validity of the geometric bound conjecture, cf.~Fig.~\ref{fig:SYK}. The bound is clearly satisfied but our inequality seems to far from tight. We have numerically verified that the ratio of $\ell_t/\ell_\lambda$ does go to 1 in the adiabatic limit but this would require to go about $10$ times slower than the data presented in~Fig.\ref{fig:SYK}. Whether the large excess energy fluctuations close to the quantum speed limit are a consequence of the numerical optimization procedure or are simply unavoidable, remains an open question.

\section{\label{eq:outro}Discussion/Outlook}

Even though standard quantum speed limit bounds are correct, they can only be saturated by a Rabi-pulse constructed out of a Gram-Schmidt orthogonalized version of the initial and target states.  
It suffices to consider an ensemble of $L$ copies of a single qubit to realize that those operations are usually not accessible in experiments. With only local controls, one can prepare a product of $L$ qubits in exactly the same time as one can prepare a single qubit state. However, currently known quantum speed limits argue that this process should be $\sqrt{L}$ times faster. This speedup is possible but requires access to a maximally-entangled state in the process. The known bounds are thus a consequence of quantum supremacy, but they do not tell us anything about how hard it is to attain the bound. In this paper, we resolved this issue by taking into account that the absolute minimal path in Hilbert space between two states cannot be attained by just any Hamiltonian; instead we compute the distance between two states as the distance in the \emph{accessible} (i.e.~fast-forward) Hamiltonian parameter space.

By reconciling ideas of Adiabatic Perturbation Theory, Counter-Diabatic driving, and Optimal Control, we conjectured that the time length $\ell_t$ for any fast-forward protocol, equal to the time integral of the instantaneous energy fluctuations, is bounded from below by the geodesic length imposed by the geometry of the instantaneous eigenstate manifold. While proving this statement for generic quantum systems remains an open problem, we have provided substantial evidence for the validity of the corresponding conjecture~\eqref{eq:conjecture}, and proved it in certain limits amenable to analytic treatment. In the exactly-solvable two- and three-level systems, we demonstrated that one can find a fast-forward Hamiltonian at the quantum speed limit analytically using ideas from counter-diabatic driving. By identifying and exploiting a residual dynamical gauge degree of freedom, we showed that the three-level system at the infinite-speed limit can be mapped to a single non-interacting collective spin degree of freedom. We also showed that {\em any} fast-forward Hamiltonian can be obtained from a counter-diabatic Hamiltonian by a unitary rotation. The mapping might allow one to prove the conjecture in general. For a non-integrable Ising chain, we used optimal control algorithms to numerically verify the universality of the proposed geometric bound not only for the ground state but also for excited states.

An interesting observation, which comes from Eq.~\eqref{eq:metric}, is that the energy fluctuations can be interpreted as the time component of the {\em non-equilibrium} quantum metric tensor $g_{tt}=\delta E_{FF}^2(t)$,
since the latter describes the distance between wave functions at two consecutive moments of time $t$ and $t+\delta t$:
\begin{eqnarray*}
	& g_{tt}=\langle \psi(t) | H_\mathrm{FF}^2(t) |\psi(t)\rangle_c= \langle \partial_t \psi | \partial_t\psi \rangle_c, \\
	& |\langle \psi(t+\delta t)|\psi(t)\rangle|^2\approx 1-g_{tt}\delta t^2.
\end{eqnarray*}
Then the conjecture~\eqref{eq:conjecture} applied to a short time interval states that, for {\em any} time evolution, the control is always time-like $g_{tt} - \dot\lambda^2 g_{\lambda\lambda}\geq 0$. The geometric bound conjecture can then be seen as a constraint imposed by causality on the optimal quantum state preparation protocols.

The universal geometric bound conjectured and checked in this paper can be used to define complexity of a dynamical control problem through the geometric length, which is a property of the ground state manifold. In particular, one can say that the problem is computationally hard if the equilibrium distance between the initial and final stated determined through the quantum geometric tensor is  exponentially large in the number of degrees of freedom. This definition of complexity makes no reference to particular protocols, which can be say realized on a quantum computer. There are very few other examples we are aware of where equilibrium properties constrain the possible behavior of a system away from equilibrium. One of them is the famous Jarzynski equality which constraints the work distribution done on a system in an arbitrary non-equilibrium process by the equilibrium free energy difference~\cite{jarzynski2011equalities}. Such results are remarkable in their nature, because they demonstrate the conservative character of physical laws, and usually point towards deeper connections between seemingly unrelated phenomena.

\begin{acknowledgments}
	\emph{Acknowledgements.---}The authors wish to thank Anatoly Dymarsky for interesting and useful discussions. MB acknowledges support from the Emergent Phenomena in Quantum Systems initiative of the Gordon and Betty Moore Foundation, the ERC synergy grant UQUAM, and the U.S. Department of Energy, Office of Science, Office of Advanced Scientific Computing Research, Quantum Algorithm Teams Program. DS acknowledges support from the FWO as post-doctoral fellow of the Research Foundation -- Flanders and CMTV. AP was supported by NSF DMR-1506340, NSF DMR-1813499, and AFOSR FA9550-16-1-0334. This research was supported in part by the National Science Foundation under Grant No. NSF PHY-1748958. We used \href{https://github.com/weinbe58/QuSpin#quspin}{QuSpin} for simulating the dynamics of the spin systems~\cite{weinberg_17,quspin2}. The authors are pleased to acknowledge that the computational work reported on in this paper was performed on the Shared Computing Cluster which is administered by \href{https://www.bu.edu/tech/support/research/}{Boston University's Research Computing Services}. The authors also acknowledge the Research Computing Services group for providing consulting support which has contributed to the results reported within this paper.
\end{acknowledgments}

\bibliography{control_RL.bib}

\newpage

\begin{widetext}
	
\newpage

\appendix

\section{\label{sec:2LS_full}Fast Forward Hamiltonian of the Two-Level System away from the Infinite-Speed Limit}

In this appendix, we show the derivation of the fast-forward Hamiltonian using counter-diabatic driving. Consider the time-dependent spin-$1/2$ Hamiltonian
\begin{equation}
H(t) = -g S^z - \lambda(t)S^x,
\end{equation}
with field coupling strengths $\lambda(t)$ and fixed $g$. We assume that $\lambda(0)=0=\lambda(T)$ and similarly for the velocity $\dot\lambda(0)=0=\dot\lambda(T)$. 

In Sec.~\ref{subsec:2LS} of the main text, we showed that the counter-diabatic Hamiltonian for this problem reads
\begin{equation}
\label{eq:H_CD_2LS}
H_\mathrm{CD}(t) = -g S^z - \lambda(t)S^x + \dot \lambda \mathcal{A}_\lambda(t),\qquad \mathcal{A}(t) = \alpha(t) S^y,
\end{equation}
where $\alpha(t)=g/[g^2+\lambda(t)^2]$ is a time-dependent strength of the gauge potential $\mathcal{A}_\lambda(t)$. Let us apply the time-dependent transformation
\begin{equation}
R(t)=\exp\left( -i\arctan\left(\frac{\dot \lambda(t)\alpha(t)}{g}\right)S^x \right).
\end{equation}
Notice how in the limit $\dot\lambda\to\infty$ one naturally obtains the step function due to the boundary condition $R(0)=1=R(T)$. This leads to the Hamiltonian
\begin{eqnarray}
H_\mathrm{FF}(t) &=& R^\dagger(t)H_\mathrm{CD}(t)R(t) - i R^\dagger(t)\partial_t R(t),\nonumber\\
&=& -g\sqrt{1 + \left(\frac{\dot \lambda\alpha}{g}\right)^2}S^z - \left(\lambda(t) + \partial_t \arctan\left(\frac{\dot \lambda\alpha}{g}\right)\right) S^x.
\end{eqnarray}
This result generalises the fast-forward Hamiltonian at the QSL, see Eq.~\eqref{eq:HFF_2LS}, which is obtained in the limiting case $\dot\lambda\to\infty$.

\section{\label{sec:controllability}Controllability of the Symmetrically-Coupled Two-Qubit Problem}

In this section, we present the details of finding the fast-forward Hamiltonian from the counter-diabatic one for the problem set discussed in Sec.~\ref{subsec:3LS_I} of the main text.

\subsection{Derivation of the Fast-Forward Hamiltonian}

The Hilbert space of symmetrically-coupled qubits Hamiltonian
\begin{equation}
H(t) = -2JS^z_1S^z_2 - g (S_1^z + S_2^z) - \lambda(t)(S_1^x + S_2^x)
\label{eq:H3LS_app}
\end{equation}
decomposes naturally into a singlet manifold, and a triplet manifold, which contains the ground states $|\psi_i\rangle$, and $|\psi_\ast\rangle$, and is preserved during the time evolution. Therefore, we can restrict the analysis to studying a three-level system (3LS) -- the simplest non-trivial generalisation of the exactly-solvable two-level system (2LS), see Sec.~\ref{subsec:2LS}. Hence, any operator on the triplet manifold is spanned by the generators of the eight-dimensional $\mathfrak{su}(3)$ algebra and the identity. The most common basis for $\mathfrak{su}(3)$ is given by the Gell-Mann matrices. However, it turns out this basis is inconvenient for our problem. Therefore, it proves useful to introduce the following basis which is more intuitive from a condensed-matter point of view:
\begin{gather}
\hat x = S_1^x + S_2^x,\quad \hat y = S_1^y + S_2^y,\quad \hat z = S_1^z + S_2^z,\nonumber\\
\hat{zz}= S^z_1S^z_2+S^z_1S^z_2,\quad \hat{xx}= S^x_1S^x_2+S^x_1S^x_2,\quad \hat{xz}= S^x_1S^z_2+S^z_1S^x_2,\quad
\hat{xy}= S^x_1S^y_2+S^y_1S^x_2,\quad \hat{yz}= S^y_1S^z_2+S^z_1S^y_2.
\end{gather}
This basis is natural for our problem, since both the Hamiltonian and the corresponding gauge potential $\mathcal{A}_\lambda$ are naturally written in basis vectors. The commutation relations between the basis vectors read
\begin{equation*}
[\hat x,\hat y]=i\hat z,\qquad [\hat y,\hat z]=i\hat x,\qquad [\hat z,\hat x]=i\hat y,
\end{equation*}
and
\begin{align*}
[\hat x,\hat{xx}]&=0,\quad &[\hat x,\hat{zz}]&=-2i \hat{yz},\quad &[\hat x,\hat{xz}]&=-i \hat{xy},\quad &[\hat x,\hat{xy}]&=i \hat{xz},\quad &[\hat x,\hat{yz}]&=i(\hat{xx}+2\hat{zz}),\nonumber\\
[\hat y,\hat{xx}]&=-2i\hat{xz},\quad &[\hat y,\hat{zz}]&=2i \hat{xz},\quad &[\hat y,\hat{xz}]&=i(\hat{xx}-\hat{zz}),\quad &[\hat y,\hat{xy}]&=-i \hat{yz},\quad &[\hat y,\hat{yz}]&=i\hat{xy},\nonumber\\
[\hat z,\hat{xx}]&=2i\hat{xy},\quad &[\hat z,\hat{zz}]&=0,\quad &[\hat z,\hat{xz}]&=i \hat{yz},\quad &[\hat z,\hat{xy}]&=-i(\hat{zz}+2\hat{xx}),\quad &[\hat{z},\hat{yz}]&=-i\hat{xz},\nonumber\\
&	  & [\hat{xx},\hat{zz}]&=0,\quad & [\hat{xx},\hat{xz}]&=-i\hat y/2,\quad & [\hat{xx},\hat{xy}]&=i\hat z/2 ,\quad & [\hat{xx},\hat{yz}]&=0,\nonumber\\
&      &	       & 		 & [\hat{zz},\hat{xz}]&=i\hat y/2,\quad & [\hat{zz},\hat{xy}]&=0,\quad 		& [\hat{zz},\hat{yz}]&=-i\hat x/2  \nonumber\\
&	  &	       &		 & 		  & 			& [\hat{xz},\hat{xy}]&=-i\hat x/4 ,\quad  & [\hat{xz},\hat{yz}]&=i\hat z/4 \nonumber\\
&	  &	       &		 &		  &			 	& 		 &				& [\hat{xy},\hat{yz}]&=-i\hat y/4  \nonumber
\end{align*}

Since the Hamiltonian~\eqref{eq:H3LS_app} is real, it can be instantaneously diagonalised by a unitary, generated by a purely imaginary operator~\cite{sels_16}. There are three independent $\mathfrak{su}(3)$ basis elements that satisfy this property which form a closed Lie subalgebra: $\mathfrak{su}(2)=\mathrm{span}\{\hat y,\hat{xy},\hat{yz}\}\subset\mathfrak{su}(3)$. Thus, in full generality, we can make the ansatz~\cite{sels_16}
\begin{equation}
\mathcal{A}_\lambda=\alpha\hat y+\beta\hat{xy}+\gamma\hat{yz},
\end{equation}
with $\alpha$, $\beta$, and $\gamma$ some $\lambda$-dependent functions [note that they are all time-dependent via $\lambda(t)$]. To determine these coefficients, it is sufficient to minimise the norm of the square of the operator $G$:
\begin{equation}
\label{eq:G}
G(\alpha,\beta,\gamma) = \partial_{\lambda}H +i [\mathcal{A_\lambda(\alpha,\beta,\gamma)},H]
\end{equation}
which is a quadratic form of $\alpha$, $\beta$ and $\gamma$. This results in the following equation
\begin{equation}
\mathrm{Hessian}\left(\|G^2(\alpha,\beta,\gamma)\|^2\right)
\left(\begin{array}{c}
\alpha \\ \beta \\ \gamma
\end{array}\right)
= -\nabla \|G^2(\alpha,\beta,\gamma)\|^2\bigg|_{\alpha=\beta=\gamma=0}
\label{eq:var_eq}
\end{equation}
for the functions $\alpha$, $\beta$, and $\gamma$. The Hessian is independent of $\alpha$, $\beta$, and $\gamma$ for a quadratic form. Finding the gauge potential via this minimization scheme  is particularly convenient as it does not require diagonalization of the Hamiltonian and returns the gauge potential in terms of expansion coefficients in the physical operator basis.

Before we proceed, we have to make a choice for the norm above. There are two natural choices -- the trace norm and the ground state norm. The former will require that every state in the initial Hamiltonian is transferred to every state in the target Hamiltonian, while the latter only enforces this for the ground state. Below, we focus exclusively on the ground state norm as in this paper we are generally interested in protocols, which target only a particular ground state.

\subsubsection{Exactly Solvable Limits}

It becomes clear that for $J=0$, when the two qubits are decoupled, the physics reduces to that of two independent two-level systems. Hence, in the limit of $J=0$, we can find the fast-forward Hamiltonian following the derivation of the fast-forward Hamiltonian in Sec.~\ref{sec:2LS_full}. We call this limiting case the 2LS limit.

Interestingly, the 3LS admits a second exactly solvable limit, $g=0$, for which the original Hamiltonian~\eqref{eq:H3LS_app} reduces to the transverse-field Ising model on two-sites. In this Ising limit, $\alpha=0=\beta$ and the gauge potential reduces to
\begin{equation}
\mathcal{A}_\lambda={\gamma}\;\hat{yz},\qquad \gamma(t) = \frac{2J^2}{4\lambda^2(t)+J^2}
\end{equation}
Notice the former similarity between this gauge potential and the one obtained for the 2LS.
It turns out, the Ising limit is another disguised two-level system, generated by the following Lie subgroup $\mathfrak{u}(2)=\mathfrak{su}(2)\oplus\mathfrak{u}(1)$, where $\mathfrak{su}(2)=\mathrm{span}
\{\hat x/2,\hat{yz},\hat{zz}+\hat{xx}/2 \}$ and $\mathfrak{u}(1)=\mathrm{span}\{(\hat{xx}-\hat{1}/3)/2\}$. As an immediate property of this decomposition, the following commutation relation follows
\begin{equation}
[\hat{xx},H(t)\big|_{g=0}]=0
\label{eq:3LS_Ising_comm}
\end{equation}

Recalling the steps we followed in the single-particle limit in Sec.~\ref{sec:2LS_full}, a rotation about the $\hat{x}$-axis should map the  gauge potential from the $\hat{yx}$ to the $\hat{zz}+\hat{xx}/2$ direction. However, there is no $\hat{xx}$ term present in the original Hamiltonian, cf.~Eq.~\eqref{eq:H3LS_app}. At first sight, the resulting rotated Hamiltonian is kicked outside the fast-forward manifold. The way out is to notice that the operator $G$, see Eq.~\eqref{eq:G}, remains invariant if we add to the gauge potential $\mathcal{A}$ any term which commutes with the Hamiltonian $H$. Thus, using Eq.~\eqref{eq:3LS_Ising_comm}, we may extend the gauge potential to 
\begin{equation}
\mathcal{A}_\lambda={\gamma}\;\hat{yz} + \rho(t)\;\hat{xx}
\end{equation}
where $\rho(t)$ is an arbitrary function of time. 

Below, we restrict to the infinite speed limit $\dot{\lambda\to\infty}$. The generalisation to arbitrary speeds can be done following the same steps as in Sec.~\ref{sec:2LS_full}. The counter-diabatic Hamiltonian in the Ising limit thus reads
\begin{equation}
H_\mathrm{CD} = \dot \lambda\left( {\gamma}\;\hat{yz} + \rho(t)\;\hat{xx} \right).
\end{equation}
To derive the corresponding fast-forward Hamiltonian, we once again do a $\pi/2$ rotation about the generator $\hat x/2$ [notice the extra factor of $1/2$ which is required by the canonical commutation relations of the emergent $\mathfrak{su}(2)$ group], and enforce the boundary condition using step functions:
\begin{equation}
R(t)=\exp\left(-i\frac{\pi}{4}\left[\Theta(t)+\Theta(T-t)\right]\hat x\right ),
\end{equation}
which leads to 
\begin{eqnarray}
H_\mathrm{FF}(t) = -\dot{\lambda(t)}\gamma(t)\left(\hat{zz}+\hat{xx}/2\right)
+ \frac{\pi}{4}\left[\delta(t)-\delta(T-t)\right] \hat x
+ \dot \lambda\rho(t)\; \hat{xx}.
\end{eqnarray}
By choosing $\rho(t)=-2\gamma(t)$, we get rid of the unwanted term to find
\begin{eqnarray}
H_\mathrm{FF}(t) = \frac{\dot \lambda(t)\gamma(t)}{J}\left(-J\;\hat{zz} +\frac{\pi}{4}\frac{J}{\dot \lambda(t)\gamma(t)}\left[\delta(t)-\delta(T-t)\right] \hat x\right)\propto H(t)\bigg|_{g=0}.
\end{eqnarray}
The quantum speed limit  in the Ising limit reads
\begin{equation}
T_\mathrm{QSL}\bigg|_{g=0}=\frac{1}{J}\int_{\lambda_i}^{\lambda_\ast}\mathrm{d}\lambda\gamma(\lambda)=
\frac{1}{J}\left(\arctan\left(\frac{J}{2\lambda_i}\right) - \arctan\left(\frac{J}{2\lambda_\ast}\right) \right).
\end{equation}

\subsubsection{General Case}

Let us now go back to the general case for the 3LS. The starting point is once again Eq.~\eqref{eq:var_eq}, with $\|\cdot\|$ the ground state norm: $\|G^2\|^2=\langle\psi_\mathrm{GS}|G^2|\psi_\mathrm{GS}\rangle$.

Curiously, choosing the ground state norm, the Hessian in Eq.~\eqref{eq:var_eq} has a vanishing determinant, which signals the existence of an additional gauge degree of freedom. Physically this freedom originates from allowing the gauge potential to mix to excited states in an arbitrary way. This freedom is encoded in choosing the operator $K$ introduced in Sec.~\ref{subsec:mapping}. Without loss of generality, we choose this along the $\hat{xy}$-direction and denote it by $b$. The reduced problem now becomes two-dimensional
\begin{equation}
\mathrm{Hessian}\left(\|G^2(\alpha,\gamma)\|^2\right)
\left(\begin{array}{c}
\alpha \\ \gamma
\end{array}\right)
= -\nabla \|G^2(\alpha,\gamma)\|^2\bigg|_{\alpha=\gamma=0}
\label{eq:var_eq_gauge}
\end{equation}
where the optimal solution $\alpha=\alpha(b)$, $\gamma=\gamma(b)$ now depends parametrically on the gauge field $b(\lambda)$. Since the exact expressions are rather cumbersome, we choose not to show them here. Instead we list the following important properties:
\begin{itemize}
	\item[(i)] the dependence on the gauge field $b$ turns out to be linear, so we can write
	\begin{equation*}
	\alpha(\lambda,b(\lambda)) = \alpha_0(\lambda) + \alpha_1(\lambda)b(\lambda),\qquad
	\gamma(\lambda,b(\lambda)) = \gamma_0(\lambda) + \gamma_1(\lambda)b(\lambda)
	\end{equation*}
	\item[(ii)] the above functions obey the following symmetries:
	\begin{equation*}
	\alpha_0(\lambda) = \alpha_0(-\lambda),\quad \gamma_0(\lambda) = \gamma_0(-\lambda),\quad \alpha_1(\lambda) = -\alpha_1(-\lambda),\quad \gamma_1(\lambda)= -\gamma_1(-\lambda).
	\end{equation*}
	\item[(iii)] for $\lambda\to 0$, both $\alpha_1(\lambda),\gamma_1(\lambda)\sim 1/\lambda$ have the same power-law divergence.
\end{itemize}

We are now fully equipped to tackle the general case. Recall that out goal is to find a time-dependent unitary $R(t)$, which maps dynamically the counter-diabatic Hamiltonian
\begin{equation}
H_\mathrm{CD}(t)=\dot \lambda\left[ \alpha(\lambda,b(\lambda))\;\hat{y} + b(\lambda)\;\hat{xy} + \gamma(\lambda,b(\lambda))\;\hat{yz} \right]
\end{equation}
to the fast-forward Hamiltonian $H_\mathrm{FF}(t)$, up to an overall time-dependent prefactor. We will construct this transformation in two steps:
\begin{itemize}
	\item[1)] Recalling that the terms in the above gauge potential form a closed $\mathfrak{su}(2)$ algebra, we use the dynamical gauge field $b(\lambda)$ to orient the gauge potential along the $\hat y$-direction. To do this, let us perform the rotation
	\begin{equation}
	R^{(1)}(t) = \exp\left(-i\arctan\left(\frac{\gamma}{2\alpha}\hat{xy}\right)\right),
	\end{equation}
	to obtain the Hamiltonian
	\begin{equation*}
	H^{(1)}_\mathrm{CD}(b(t),t) = \dot \lambda\left[ \frac{1}{2}\sqrt{4\alpha^2+\gamma^2}\;\hat{y} + \left(b - \frac{2}{\dot \lambda}\partial_t \arctan\left( \frac{\gamma}{2\alpha} \right) \right)\;\hat{xy} \right].
	\end{equation*}
	Clearly, we can eliminate the $\hat{xy}$ term, provided the gauge field satisfies the following nonlinear first-order differential equation:
	\begin{equation}
	\label{eq:b_ODE_app}
	b(\lambda) = 2\partial_{\lambda}\arctan\left( \frac{\gamma(\lambda,b(\lambda))}{2\alpha(\lambda,b(\lambda))}\right),\qquad \gamma(\lambda_i,b(\lambda_i))=0=\gamma(\lambda_\ast,b(\lambda_\ast)),
	\end{equation}
	where the boundary conditions (BC) are chosen to satisfy the requirement $R^{(1)}(0)=\hat 1=R^{(1)}(T)$. This is at first sight problematic, since we have two BC for a single first-order ODE. However, from the symmetry properties above, one can convince oneself that if $b(\lambda)=-b(-\lambda)$ is antisymmetric, then both BC coincide and thus represent a single constraint, since $\lambda_\ast=-\lambda_i$. Indeed, using the same symmetry properties, it is easy to see that $b(-\lambda)$ also obeys Eq.~\eqref{eq:b_ODE_app}. Last, notice that, even though the functions $\alpha$ and $\gamma$ have a $1/\lambda$-singularity for $\lambda\to 0$, the quotient $\gamma/\alpha$ does not, as the singularity is lifted. Thus, the RHS of Eq.~\eqref{eq:b_ODE_app} is a smooth function of $b$ and $\lambda$. It then follows from the Picard-Lindel\"of theorem for existence and uniqueness of ordinary differential equations that the initial value problem in Eq.~\eqref{eq:b_ODE_app} has a unique solution.
	
	In the following, let us fix the dynamical gauge field $b$ to satisfy Eq.~\eqref{eq:b_ODE_app}. Then the counter-diabatic Hamiltonian after the first rotation reads
	\begin{equation*}
	H^{(1)}_\mathrm{CD}(t) =  \frac{\dot \lambda}{2}\sqrt{4\alpha^2+\gamma^2}\;\hat{y}.
	\end{equation*}
	
	\item[2)] We can now perform the $\hat x$-rotation, familiar from the single-particle limit :
	\begin{equation}
	R^{(2)}(t) = \exp\left(-i\frac{\pi}{2}\left[\Theta(t)+\Theta(T-t)\right]\hat{x}\right),
	\end{equation}
	which satisfies the BC $R^{(2)}(0)=\hat 1=R^{(2)}(T)$. This transforms the counter-diabatic Hamiltonian to
	\begin{equation*}
	H_\mathrm{FF}(t) =  -\frac{\dot \lambda}{2}\sqrt{4\alpha^2+\gamma^2}\;\hat{z} + \frac{\pi}{2}\left[\delta(t)+\delta(T-t)\right]\;\hat{x}
	\end{equation*}
	which is precisely the fast-forward Hamiltonian we used in the main text.

\end{itemize}

\end{widetext}

\end{document}